**Atomic and electronic structure of doped Si(111)(2√3×2√3)R30º-Sn interfaces**

Seho Yi,[1,*] Fangfei Ming,[2,*] Ying-Tzu Huang,[3] Tyler S. Smith,[2] Xiyou Peng,[2] Weisong Tu,[2] Daniel Mulugeta,[2] Renee D. Diehl,[3] Paul C. Snijders,[4,2] Jun-Hyung Cho[1,5] and Hanno H. Weitering[2]

[1] *Department of Physics and Research Institute for Natural Sciences, Hanyang University, Seoul 133-791, Korea*

[2] *Department of Physics and Astronomy, The University of Tennessee, Knoxville, Tennessee 37996*

[3] *Department of Physics, Penn State University, University Park, Penn State University, Pennsylvania 16802*

[4] *Materials Science and Technology Division, Oak Ridge National Laboratory, Oak Ridge, TN 37831*

[5] *ICQD, Hefei National Laboratory for Physical Sciences at the Microscale, and Synergetic Innovation Center of Quantum Information and Quantum Physics, University of Science and Technology of China, Hefei, Anhui 230026, China*

*) These authors contributed equally

**ABSTRACT**

The hole doped Si(111)(2√3×2√3)R30º-Sn interface exhibits a symmetry-breaking insulator-insulator transition below 100 K that appears to be triggered by electron tunneling into the empty surface-state bands. No such transition is seen in electron-doped systems. To elucidate the nature and driving force of this phenomenon, the structure of the interface must be resolved. Here we report on an extensive experimental and theoretical study, including scanning tunneling microscopy and spectroscopy (STM/STS), dynamical low-energy electron diffraction (LEED) analysis, and density functional theory (DFT) calculations, to elucidate the structure of this interface. We consider six different structure models, three of which have been proposed before, and conclude that only two of them can account for the majority of experimental data. One of them is the model according to Törnevik et al. [C. Törnevik et al., *Phys. Rev. B* **44**, 13144 (1991)] with a total Sn coverage of 14/12 monolayers (ML). The other is the 'revised trimer model' with a total Sn coverage of 13/12 ML, introduced in this work. These two models are very difficult to discriminate on the basis of DFT or LEED alone, but STS data clearly point toward the Törnevik model as the most viable candidate among the models considered here. The STS data also provide additional insights regarding the electron-injection-driven phase transformation as well as the critical role of valence band holes in this process. Similar processes may occur at other metal/semiconductor interfaces, provided they are non-metallic and can be doped. This could open up a new pathway toward the creation of novel surface phases with potentially very interesting and desirable electronic properties.

## I  INTRODUCTION

Surfaces and ultrathin metal/semiconductor interfaces are an interesting platform for studying phase transitions and emergent phenomena in two-dimensions (2D). In particular, adsorption of group III and group IV post-transition metal atoms on Si(111) and Ge(111) surfaces produces a variety of interesting phenomena such as Mott metal-insulator transitions [1-3], charge-ordering transitions [4-8], and even superconductivity [9-13]. Although the structural degrees of freedom of these systems are determined in large part by the underlying substrate, these systems are near-perfect 2D electron systems as the electronic interactions take place in one or several surface state bands that are generally decoupled from the 3D electronic structure of the underlying Si or Ge substrate.

Notwithstanding their intellectual appeal, the electronic properties of surfaces and interfaces are difficult to control, other than through simple coverage control or change of substrate. For instance, strictly 2D systems such as exfoliated 2D materials, surfaces, and interfaces are all difficult to dope because the presence of dopants inevitably introduces structural disorder as the ionized dopant impurities, and the associated modulations of the potential energy landscape, become an integral part of the 2D electron system. In principle, this dilemma can be avoided by employing a modulation doping approach in which the chemical dopants are spatially separated from the 2D electron system [14], as is done in, *e.g.*, semiconductor quantum well superlattices [14, 15] and layered perovskite materials [16]. In fact, most of the current emphasis on low-dimensional quantum matter phases involves mapping of the electronic phase diagrams of quasi 2D bulk materials as a function of doping level or chemical potential. However, efforts to systematically control the electronic properties of surfaces and interfaces are largely undeveloped, suggesting that many interesting 2D phases of matter are still awaiting experimental discovery.

Using the modulation doping concept, we recently uncovered a novel equilibrium phase in a hole-doped bilayer of Sn on p-type Si(111) [17]. The holes originate from the boron dopants inside the bulk substrate. The formation of this broken symmetry phase appears to be triggered by electrons tunneling from the tip of a scanning tunneling microscope (STM) into the sample, and sets in below 100 K. No such transition is seen on n-type Si suggesting that the high-symmetry phase is the ground state for the n-type system.

Scanning Tunneling Microscopy (STM) images of the high-symmetry phase or Si(111)($2\sqrt{3}\times2\sqrt{3}$)$R30^\circ$-Sn surface (henceforth denoted as the '$2\sqrt{3}$ phase') reveal a 2D hexagonal array of Sn tetramers where each tetramer consists of a bright 'up-dimer' and a dim 'down dimer' (Fig. 1; see also Ref [17]). The $2\sqrt{3}$ surface has a rhombic unit cell with *cm* symmetry (2D space group no. 5) [17] where the ($2\sqrt{3}\times2\sqrt{3}$) supercell is rotated 30° degrees with respect to the (1×1) unit cell of the truncated Si(111) substrate. The broken symmetry phase, or Si(111)($4\sqrt{3}\times2\sqrt{3}$)R30°-Sn surface ('$4\sqrt{3}$ phase'), reveals a pattern in which the bright dimers are rotated 45° and form a staggered zig-zag pattern (Fig. 1). The easiest way to picture this structural transition is to consider a centered rectangular lattice for the high symmetry phase, and position the Sn tetramers at the lattice points. The staggered arrangement of the dimers in the $4\sqrt{3}$ phase then implies a loss of the centering translation, resulting in a rectangular unit cell with *pg* symmetry (2D space group no. 4). Interestingly, the transition is not driven by a Fermi surface instability as the $2\sqrt{3}$-Sn is insulating with a band gap of approximately 0.45 eV (at 77 K). Furthermore, the transition is both displacive and ferroic in nature [Ref. 17 including Supplementary text], which makes this system unique among known metal-semiconductor interfaces.

To understand the driving force of this symmetry-breaking transition in the hole-doped system and particularly the role of dopants, valence band holes, and current injection in the transition, the structure of this interface must first be resolved. This has proven to be a daunting task because the coverage is not precisely known. The $2\sqrt{3}$ interface is known to consist of a double layer of Sn on Si(111) where the top layer is comprised of four atoms (one tetramer) per $(2\sqrt{3} \times 2\sqrt{3})R30^{\circ}$ unit cell [17-24]. The unit cell is furthermore believed to contain a total of 13 or 14 Sn atoms [18-23], implying a coverage of 13/12 or 14/12 monolayers (ML), although some of the most recent models such as the ones by Eriksson [24] and by Srour et al. [25] suggest a lower coverage (1 ML = 7.84 $\times 10^{14}$ atoms per cm$^2$ which is the areal density of Si atoms in the Si(111) plane). This structure is particularly difficult to solve because the second Sn layer is mostly invisible to the STM whereas the large number of atoms per unit cell makes it very difficult to discriminate between models with say 12, 13, or 14 atoms per unit cell with the available surface analytical techniques.

In this paper, we will present a combined experimental and theoretical study in an attempt to resolve the atomic structure of the $2\sqrt{3}$ structure. This paper is organized as follows. Section II provides a brief description of the experimental and theoretical methods used in this work. In Section III, we will narrow down the coverage of the $2\sqrt{3}$ phase using STM and x-ray photoelectron spectroscopy (XPS). In the following sections, Sections IV and V, we employ density functional theory (DFT) and low-energy-electron diffraction (LEED) to explore the viability of the remaining structure models as well as that of several new models with similar coverage. Specifically, we will use the total-energy-minimized coordinates from the DFT calculations as input for the structure optimization procedure in LEED where the experimental LEED I-V data are fitted against dynamical LEED calculations. On the basis of total energy considerations from DFT and the overall Pendry R-factor [26] from LEED, the likely structure can be narrowed down to two candidate models, the one introduced by Törnevik et al. [18] with 14/12 ML coverage, and the 'revised trimer model' with 13/12 ML, introduced in this work.

Following a brief discussion of the low-temperature 4√3 phase in Section VI, we present new STS spectra of the 2√3Sn and 4√3Sn surface in Section VII. The experimental LDOS will be compared with the computed DOS for the revised trimer and Törnevik models, and it is concluded that the Törnevik model [18] best accounts for the experimental data. In the last section of this paper, Section VIII, we offer a speculative scenario regarding the role of the tunneling electrons and valence band holes in the structural transition from the high-temperature 2√3Sn to the low-temperature 4√3Sn configuration, and vice-versa. Our study indicates that the transition is precipitated by tunneling into 'anti-bonding' dangling bond surface states that are localized on the 'down atoms' of the Sn tetramer. At 4.4 K, the process must involve quantum tunneling of the Sn adatoms between different up-down configurations. The tunneling barrier for this transition is strongly reduced by hole doping, which explains why the 4√3Sn structure is only observed on the p-type substrates. This concept might be applicable to other doped surface systems, provided they are gapped, and possibly provides a pathway toward the realization of other novel surface phases.

## II    EXPERIMENTAL AND THEORETICAL PROCEDURES

The sample preparation procedures in this paper are similar to those reported in Ref. 17. Briefly, the Si(111)(2√3×2√3)$R$30°-Sn surface reconstruction was prepared in ultrahigh vacuum by evaporating approximately 1 ML of Sn from a conventional effusion cell with BN crucible and PID temperature controller onto thoroughly degassed n-type Si(111)7×7, p-type Si(111)7×7, and Si(111)(√3×√3)$R$30°-B substrates. (Details concerning the coverage calibration will be presented in Section III). These substrates have a room temperature resistivity of 0.002 Ω•cm, 0.03 Ω•cm, and 0.001 Ω•cm, respectively [27, 28]. The resulting structures will be denoted as n-2√3Sn, p-2√3Sn, and B-2√3Sn, respectively. The Si(111)($\sqrt{3}\times\sqrt{3}$)$R$30°-B substrate is heavily doped and contains a 2D ordered array of substitutional boron dopants in the second silicon layer [27, 28]. The substrate temperature was kept at 550 °C during the deposition and post-annealing at 550 °C for several more minutes.

After cooling to room temperature, samples were transported in UHV to a neighboring UHV chamber for *in-situ* surface characterization with variable-temperature STM, LEED I-V, and X-ray photoelectron spectroscopy (XPS). The variable temperature STM allows for atomic resolution imaging and spectroscopy for sample temperatures ranging from 40 K to 300 K. Additional STM experiments were conducted in a separate low-temperature STM system that reaches a temperature of 4.2 K. Chemically etched tungsten tips were used. The dI/dV signal was acquired by a lock-in amplifier using a typical modulation frequency of 831 Hz and a modulation amplitude of ~ 10 mV.

XPS measurements were performed using a monochromatized X-ray source producing Al K$_\alpha$ radiation ($h\nu$ = 1486.6 eV). Photoelectrons were dispersed using a 125 mm radius electron energy analyzer and counted using a multi-channel detector. Core levels were measured with low angular resolution (±8°) so as to minimize photoelectron diffraction effects in the determination of the Sn coverage. XPS was also used to determine the band bending below the surface, as described in detail in Ref. [17]. The overall resolution of the XPS instrument is about 0.3 eV at 25 eV pass energy.

LEED patterns were recorded using a rear-view LEED system and CCD camera. The LEED spot intensities were extracted using the in-house EasyLEED software [29, 30], which automatically tracks the same beam spot across successive LEED frames. The LEED pattern exhibits mirror plane and three-fold rotational symmetry, reflecting the domain averaging and substrate symmetry. These intensities were symmetry averaged to obtain 20 independent beams. The total energy range of these 20 beams is 5120 eV and 7460 eV for the n-2$\sqrt{3}$Sn and B-2$\sqrt{3}$Sn structures, respectively. The structure analysis was carried out using automated tensor LEED software (SATLEED) [31]. The atomic scattering phase shifts ($lmax$=8) for the two Sn layers, the surface Si layer, the deeper Si layers, and the bulk Si layer were calculated using the Elastic Electron-Atom Scattering in Solids and Solid Surfaces (EEASiSSS) program [32]. In the structural analysis, the imaginary part of the inner potential was fixed at -4.5 eV whereas the real part of the inner potential was optimized by fitting the calculated I(V) spectra to the experimental data. The atomic coordinates in Sn layer and the first three Si bilayers were adjusted along the direction perpendicular to the surface, using an automated search procedure. Next, we performed full 3D structure optimizations for the Sn layers, and optimized the corresponding Debye temperatures. The agreement between the experimental and calculated I-V curves was quantified using the Pendry R-factor [26].

DFT calculations of the various structure models were performed using the FHI-aims [33] code for an accurate, all-electron description based on numeric atom-centered orbitals, with "tight" computational settings. For the treatment of the exchange-correlation energy, we employed the generalized-gradient approximation functional of Perdew-Burke-Ernzerhof (PBE) [34] and the hybrid functional of Heyd-Scuseria-Ernzerhof (HSE) [35]. The system was modeled by a periodic slab geometry with 30 Å of vacuum in between the slabs. Each slab consists of four silicon bilayers plus a Sn layer. The Si atoms in the bottom layer of the slab are passivated with hydrogen. We employed a dipole correction that cancels the artificial electric field across the slab [36]. The k-space integrations in the $2\sqrt{3}$ and $4\sqrt{3}$ unit-cell were done with 144 and 72 k points, respectively. All atoms were allowed to relax along the calculated forces until all the residual force components were less than 0.02 eV/Å.

Finally, STM images of the various structure models were simulated using the Tersoff-Hamann approximation [37, 38]. The simulated filled-state and empty state images were obtained by integrating the charge density from the Fermi level up to the voltages indicated, at ~3 Å above the topmost surface atom.

### III COVERAGE DETERMINATION

Most literature data point to a coverage of about 1 ML for the Si(111)($2\sqrt{3}\times2\sqrt{3}$)$R$30°-Sn interface. The most direct determination of the *absolute* coverage was done with ex-situ Rutherford Backscattering Spectrometry, which produced a number close to $1.2 \pm 0.1$ ML [19]. This is consistent with a structure model containing 13 or 14 atoms per $(2\sqrt{3}\times2\sqrt{3})R30°$ unit cell. The implicit assumption in this estimate is that the surface is homogeneously covered with the $2\sqrt{3}$ phase and that that no material is lost while transporting the sample through air [19].

Here, we use in-situ probes to provide another coverage estimate for the $2\sqrt{3}$ phase on both *n*-type and *p*-type substrates. First, we optimized the sample preparation procedure to cover the surface uniformly with the low-density $Si(111)(\sqrt{3}\times\sqrt{3})R30°$-Sn (or $\sqrt{3}$) phase, whose coverage is known to be exactly 1/3 ML [39]. The uniformity of the surface was checked with STM. Apart from the presence of the usual surface defects, we were able to achieve almost uniform $\sqrt{3}$ coverage on n-type and lightly-doped *p*-type substrates, and hence this surface provides a suitable calibration for XPS coverage determination. The Sn coverage on more heavily doped *p*-type substrates becomes increasingly non-uniform and hence coverage estimates for those samples are less reliable.

As a next step, we recorded the Si $2p$ and the Sn $3p$, $3d$, and $4d$ core level spectra for three different escape angles in-situ with XPS and subsequently normalized the Sn core level intensities to that of the Si $2p$ spectrum (see Appendix). Next, we repeated this procedure for the $Si(111)(2\sqrt{3}\times2\sqrt{3})R30°$-Sn surface with the unknown coverage and compared the normalized core level intensities of the $2\sqrt{3}$ and $\sqrt{3}$ phases, as given in Table I. For further details, we refer to the Appendix. We conclude, ignoring possible photoelectron diffraction effects, that the coverage of the $2\sqrt{3}$ phase is $1.05 \pm 0.13$ ML where the error margin is the standard deviation for the measurements recorded from the different escape angles and core levels. (Excluding the two outliers in Table 1, the total coverage would average to 1.11 ML). Intensity variations due to possible photoelectron diffraction depend on the precise placement of the Sn atoms, which we consider to be unknown at this point, as well as the electron kinetic energy and detector angle. A detailed photoelectron diffraction study is beyond the scope of this work, although it should be noted that our margin of error incorporates measurements with different photoelectron kinetic energies and different escape angles, using low angular resolution.

This initial coverage estimate is in quite good agreement with the RBS measurement by Törnevik at al. [19], as well as other estimates by e.g. Ottaviano et al. [22]. It is consistent with structural models containing 13 or 14 Sn atoms per $2\sqrt{3}$ unit cell, such as the models by Ichikawa et al. [21] and Törnevik et al. [19]. The 1.0 ML model by Eriksson et al. [24] is fully consistent with this coverage estimate but the most recently proposed structure model by Srour et al. [25] with 0.5 ML of Sn is well outside this margin of error.

In a second experiment, we exposed the first Sn layer of the $2\sqrt{3}$Sn reconstruction by applying a voltage pulse to the STM tip. Fig. 2(a) shows the exposed surface area surrounded by domains of the $\sqrt{3}$ and $2\sqrt{3}$ phases. The Sn tetramers on top of the first layer have disappeared. The bright feature in the middle of the image likely represents the pile-up of the original tetramer atoms. The exposed area is labeled as a Sn $2\sqrt{3}$* phase as it also features a $(2\sqrt{3}\times2\sqrt{3})R30°$ unit cell, although with a different structure than the $2\sqrt{3}$ cell observed in the top layer. Panels 2(b) and 2(c) are close-up views of the $2\sqrt{3}$* structure at different scanning biases. The +0.1 V image shows two trimer units and a central bright feature within the $2\sqrt{3}$* unit cell. These bright features are clearly resolved in the +1.3 V image, and consist of a trimer unit that is rotated 180° relative to the trimers in the +0.1 V image. This structure is clearly metastable as it can only be accessed by pulsing the voltage on the STM tip. This lower level structure was observed on all substrates with pulsing, but on the n-type substrate it requires a much higher voltage pulse of over 5 V, indicating that the n-$2\sqrt{3}$ structure is relatively more stable than the p-$2\sqrt{3}$ and B-$2\sqrt{3}$ interfaces. Although one cannot be sure how many tin atoms were stripped from the original $2\sqrt{3}$ structure, these results suggest that the underlayer contains nine tin atoms per $2\sqrt{3}\times2\sqrt{3}$ unit cell, forming two 'up' and one 'down' pointing trimer once the tetramer of the top Sn layer is removed, implying a total coverage of 13 atoms per $2\sqrt{3}$ unit cell (1.08 ML).

## IV    STRUCTURE MODELS FROM DENSITY FUNCTIONAL THEORY

Inspired by the coverage estimates from the previous section, and particularly the atomic resolution images of the exposed underlayer, which revealed the existence of three trimer units, we focus our attention to structure models containing 13 or 14 atoms per unit cell. The Eriksson model with 12 atoms per unit cell was found to be highly unstable. Figure 3 shows total energy minimized structures of the various structure models considered in this paper. Each panel contains four $2\sqrt{3}$ unit cells with one unit cell indicated by the shaded area. Gray atoms are the silicon atoms of the substrate. Light blue atoms are the Sn atoms in the first layer and the dark blue atoms represent the Sn tetramers seen in the STM images. The inset shows the side view of the structure, as seen along the long diagonal of the supercell.

*Törnevik model*. The total-energy-minimized Törnevik structure in Fig. 3(a) contains ten Sn atoms plus the Sn tetramer seen in STM. The side view image indicates that the tetramer is strongly buckled where the 'up atoms' seen in STM protrude about 1.5 Å above the rest, consistent with the bilayer structure seen in STM. The up atoms are approximately located above the $T_4$ adatom sites of the Si substrate. The down atoms are located above the three-fold hollow ($H_3$) sites. This registry is consistent with the one reported by Törnevik et al. [18], which is confirmed in this work (Fig 4). Eight Sn atoms are four-fold coordinated in a distorted tetrahedral bonding configuration. The tetramer atoms as well as the two atoms that are located on the short diagonal of the rhombic unit cell have one dangling bond each. These six dangling bonds are nominally half-filled but the surface structure relaxes so as to create a semiconducting ground state with a 0.55 eV band gap. For a detailed discussion of the electronic structure, we refer to Section VII.

*Revised Törnevik model.* This model in Fig. 3(b) is essentially the same as the Törnevik model, except for the large lateral displacement of the two Sn atoms on the short diagonal. This structure is very stable. Both dimers are in the 'up' position, which seems inconsistent with the STM experiment.

*Trimer model*.    The trimer model was inspired by the adatom stripping experiment in Section III (Fig. 2). Here, we placed three trimer units in the $2\sqrt{3}$ unit cell as suggested by the trimer registry of the exposed underlayer (Fig. 2(d)), and place four additional Sn atoms on top. The fully relaxed structure is shown in Fig. 3(c). Only the central trimer unit remains recognizable as such. The other two trimers are heavily distorted as the Sn atoms tend to form chain-like structures. This chain-like backbone structure looks very similar to that of the Törnevik model. In fact, the only major difference between these two models is the missing Sn atom that was located on the short diagonal near the up atoms of the Törnevik structure. Other noticeable differences with the Törnevik model are that the structure is almost flat, *i.e.*, it has lost the characteristic bilayer structure suggested by STM, and that the bond lengths of the two dimers within the Sn tetramer have become very different. The dangling bond count is still the same, with one dangling bond now located on the Si atom below the short dimer. The absence of a bilayer configuration suggests that this model should be ruled out.

*Revised trimer model.* This model differs from the trimer model in that instead of removing the Sn atom located on the short diagonal and bonded to the up dimer of the Törnevik model, we now remove the other atom on the short diagonal, i.e., the one bonded to the down dimer (Fig. 3(d)). Unlike the trimer model, this structure retains its bilayer character and maintains the height difference between the two dimers, as in the Törnevik model. The dangling bond and electron count is the same as the trimer model.

*Ichikawa model.* The model proposed by Ichikawa and Cho [21] was motivated by the apparent discrepancy between the Törnevik model [18, 19] and published Patterson function from surface x-ray diffraction data [20]. Using their optimized DFT coordinates, we obtained the structure shown in Fig. 3(e). This structure is quite different from the previous ones, which all contained a similar back bone structure of the underlayer. This structure also seems inconsistent with the bilayer geometry seen in STM.

*Formation Energy*:   To compare the total energies of the above structure models, one needs to take into account the fact that the 13/12 ML and 14/12 ML models differ in coverage. Here, we compare the formation energies $E_f$ as follows:

$$\Delta E_f = E_{model} - \Delta N \mu_{Sn} - E_{Tornevik} \tag{1}$$

where the formation energy is considered relative to that of the Törnevik model. $E_{model}$ is the total energy of a given structure model calculated from DFT using the PBE functional. The total energy scales with the number of Sn atoms in the slab, which can be corrected for by taking into account the chemical potential of the Sn atoms, $\mu_{Sn}$ [40]. The upper limit of $\mu_{Sn}$ corresponds to the chemical potential of bulk α-Sn. The lower limit depends on the experimental conditions, as will be discussed in a moment, and is taken as a free parameter for the purpose of the calculations. Since the Törnevik model contains 14 atoms per unit cell, $\Delta N$ = -1 or 0 for models containing 13 or 14 atoms per unit cell, respectively.

The formation energies of the various $2\sqrt{3}$ structure can now be compared in a meaningful way and the result is shown in Fig. 5. The graph indicates that the revised Törnevik and revised trimer models are thermodynamically most preferred. Note, however, that the Törnevik and revised Törnevik models are very close in energy and only differ by 0.13 eV per $2\sqrt{3}$ unit cell, which is less than 10 meV per Sn atom. The Törnevik model should therefore not be ruled out. Ichikawa's model and the trimer model are least preferred, regardless of the chemical potential. The revised trimer model is most preferred among the models considered for $\mu_{Sn}$ < -0.4 eV. The experimental value of the chemical potential in the formation of the $2\sqrt{3}$ phase can be estimated from

$$\mu_{Sn} = \mu_{\alpha Sn} + k_B T \ln\left(p_{cell} \,/\, p_{sample}\right) \qquad (2)$$

assuming quasi equilibrium growth [40]. Here, $p_{cell}$ is the vapor pressure of Sn inside the effusion cell, $p_{sample}$ the vapor pressure at the sample, and $T$ the sample temperature. Accordingly, one finds that $\mu_{Sn}$ is about -0.85 eV which leaves the revised trimer model as the most probable structure on the basis of total energy considerations within DFT.

Finally, we calculated the total energy minimized structures of the revised trimer and Törnevik models in the presence of a $(\sqrt{3}\times\sqrt{3})R30°$-B underlayer where the boron atoms are located at the $S_5$ lattice location [27, 28] right below the $T_4$ adatom sites of the Si substrate (Fig. 3(f) and 3(g)). There are three possible ways to distribute the boron atoms on the $T_4$ sublattice. In the calculation, we only considered the boron distribution that preserved the *cm* space group symmetry of the 2√3 structure with four boron atoms per 2√3 unit cell. The fully relaxed structures appear almost perfectly identical to those without boron underneath, which is fully consistent with our previous report stating that the 2√3 structures on the n-type and heavily doped p-type systems are identical.

## V  STRUCTURE DETERMINATION FROM LEED I(V)

LEED patterns of the B-2√3Sn and n-2√3Sn surfaces are shown in Fig. 6. The I(V) data of the n-2√3Sn surface are shown in Fig. 7. The latter are fitted against the calculated LEED intensities for the structure models considered in Fig. 3. We used the optimized DFT coordinates of the (revised) Törnevik and (revised) trimer models as starting point of the structure optimization for the n-2√3Sn and B-2√3Sn systems.

The overall Pendry R-factors for both systems are tabulated in Table 2. Evidently, the trimer model agrees best with the LEED data for the n-2√3Sn system, even though it is one of the least favorable models according to the total energy calculation in DFT.   (The trimer model also seems inconsistent with the observed double-layer structure in STM). The R-factors of the Törnevik and revised trimer models for the n-2√3Sn system are very close, essentially indistinguishable, and the revised Törnevik model is clearly the least favorable. Attempts to find a new stable structures in the DFT, starting from the LEED optimized coordinates, were unsuccessful as the atoms relaxed back to the original positions found with DFT.

Note that the Pendry R-factors of the B-2√3Sn structure are substantially higher, though it should be kept in mind that the data sets are also larger. Here, the trimer and Törnevik models are in closest agreement with the experimental data. To put the higher R-factors for the B-2√3Sn models into perspective, we directly compared the experimental I(V) data of the n-2√3Sn and B-2√3Sn surfaces and find that the mutual Rp factor is 0.27. Hence, experimentally the structures of these two systems seem reasonably close. This observation is strongly corroborated by the STS tunneling spectra in Section VII.

It is clear that the LEED I-V data have difficulty discriminating between the various structure models,

although arguably one could eliminate the revised Törnevik model as the least likely candidate. In restrospect, this difficulty may not be very surprising as the number of atoms per unit cell is large and the differences between the structure models are quite subtle. The relatively large R-factors nonetheless suggest that we may not have found the right structure. However, as we will show in Section VII, among the models considered here, the Törnevik model is in best agreement with our STM and STS data.

For completeness, we tested the validity of the 0.5 ML structure proposed by Srour et al. [25], and found an overall R-factor of 0.64, using their DFT coordinates and 0.58 after further optimization with the SATLEED code. This model can thus be ruled out as a viable candidate.

## VI    THE LOW TEMPERATURE PHASE

In order to find the structure of the low temperature $4\sqrt{3}$Sn phase, we employed a $(4\sqrt{3}\times2\sqrt{3})R30°$ supercell. Calculations starting with the Törnevik coordinates produced a meta-stable $4\sqrt{3}$Sn phase with a total energy of +540 meV and +400 meV per unit cell relative to that of the $2\sqrt{3}$Sn and B-$2\sqrt{3}$Sn phase, respectively (Fig. 8). A similar metastable structure was found for the revised trimer model. Its total energy is 300 meV higher than that of the corresponding $2\sqrt{3}$ phase. Simulated STM images for these distorted $4\sqrt{3}$ structures are also shown in Fig. 8. It is clear that the Törnevik model provides a better match in that the intensity of the two dim atoms within each tetramer seems equivalent, consistent with experiment, while in the revised trimer model one of the two dim features appears to be missing in the simulation.

It is still too early to proclaim the Törnevik model as the model that appears most consistent with experiment. First of all, one cannot be sure that the $4\sqrt{3}$ structure found by DFT is the one observed in experiment, especially since its total energy is so much higher than that of the $2\sqrt{3}$Sn phase. On the other hand, one recalls that the $4\sqrt{3}$ does not form spontaneously and that it can only be formed during empty state imaging, meaning that current is injected into the sample. Hence, it could be a meta-stable state. We also investigated the role of a possible electric field effect on the relative stability of the $2\sqrt{3}$ and $4\sqrt{3}$ phase, but found this effect to be rather small. Technical details of this procedure can be found in Ref. [41]. The most troubling aspect is that theoretically, the band gap of the metastable 4√3 structure is smaller than that of the 2√3 phase, while experimentally the 4√3 phase has the larger band gap. Hence, it is likely that the 4√3 phase would be the most stable state, contrary to the DFT prediction. This calls the validity of the found 4√3Sn structure into question.

## VII    ELECTRONIC STRUCTURES

Fig. 9 shows the STM image simulations for the total-energy minimized structures of the various structure models discussed in this paper. The presence or absence of boron in the second silicon layer does not affect the visual appearance of these simulated images. As anticipated in Section IV, the revised Törnevik model as well as the trimer and Ichikawa models do not reproduce the stark contrast between the 'up-dimer' and 'down-dimer' in the experimental STM images, as their Sn layers are almost flat. The agreement is poor regardless of the tunneling bias. On the other hand, the simulated images for the Törnevik and revised trimer models closely resemble the experimental images. Among the models considered in this paper, the Törnevik model and revised trimer model are the only viable candidates still remaining.

To further discriminate these two models, we consider their electronic structures, and particularly compare the local density of states obtained from the tunneling dI/dV spectra, measured at 4.4 K, and the DFT results. To this end, we implemented a separate set of DFT calculations using the HSE functional [35]. The hybrid exchange functional produces band gaps that are more accurate than those obtained with the (cheaper) PBE calculations, which tend to underestimate the band gaps. However, before comparing the theoretical and experimental results, key aspects of the experimental dI/dV data will be discussed first.

In Ref. 17, it was shown that the dI/dV spectra of the n-type and p-type 2√3Sn surfaces, including the B-2√3Sn surface, were very similar, except for the rigid 0.36 eV shift of the chemical potential from 0.03 eV above the conduction band minimum for n-2√3Sn to 0.12 eV above the valence band maximum for the

p-2√3Sn and B-2√3Sn systems. These spectra were acquired at 77 K and the band gap was 0.45 eV. As pointed out in Ref. 17, the presence of a boron (√3×√3)$R$30° underlayer prior to the deposition of Sn does not leave any boron-induced signature in the dI/dV data or LDOS (77 K). This will be an important point in our effort to reconcile the results from STS, LEED, and DFT, to be discussed below.

Here we present new STS data, this time the data are acquired at 4.4 K. Due to the high resistivity of the n-type material at 4.4 K, we could only measure dI/dV spectra reliably for the hole-doped systems [42, 17]. However, in light of the 77 K data mentioned above, we conjecture that the dI/dV spectra of the B-2√3Sn system reflect the LDOS of the n-2√3Sn, p-2√3Sn, and B-2√3Sn systems. Only the latter is experimentally accessible at 4.4 K. The B-2√3Sn system undergoes a transition to the 4√3Sn phase when electrons are being injected into the surface. To avoid the complications due to the structural change induced by the tunnel current, we only recorded STS data from structurally stable 2√3Sn and 4√3Sn domains. This is possible because at 4.4 K, about 20% of the surface remains covered by stable 2√3Sn domains [17].

Figs. 10(a) and 10(b) shows the normalized B-2√3Sn spectra recorded at 77 K and 4.4 K, respectively. The spectra are single-point spectra recorded on the up-atom location. Normalized means that the dI/dV spectra are divided by the simultaneously acquired I(V) spectrum to correct for the exponential voltage dependence of the tunneling current. This procedure provides a more accurate representation of the LDOS [43]. Each d$I$/d$V$/($I$/$V$) spectrum is obtained by dividing the d$I$/d$V$ curve by its smoothed $I$($V$) curves, such that dividing zero is avoided within the gap region [44]. Note that the spectral features of the 77 K and 4.4 K spectra line up quite nicely, although the spectral features of the 4.4 K data are much sharper.

In Fig. 11(a) and 11(b), we compare the normalized dI/dV spectra of the B-2√3Sn and B-4√3Sn phases. These are single-point spectra recorded on a down-atom. The band gaps are 0.45 and 0.70 eV, respectively, which strongly indicates that the B-4√3Sn structure is lowest in energy. Fig 11(c) demonstrates an almost perfect line up of the spectral features of the 2√3Sn and 4√3Sn phases, after accounting for the difference in band gap, except for the missing peak at 1.7 V in the B-4√3Sn spectrum. As we will show in Fig. 12, the B-4√3Sn spectrum does in fact contain a peak at 1.7 V when recorded on the up-atom location. While there is no 'a priori' reason why the spectra of the 2√3Sn and 4√3Sn structures should be lined-up this way, such precise matching of the peak positions is probably not coincidental. Such a line-up should be reproduced in the theoretical DOS of the 2√3Sn and 4√3Sn structures.

Fig. 12 shows single-point spectra recorded at different locations within neighboring B-2√3Sn (left) and B-4√3Sn (right) domains. In particular, the B-2√3Sn structure shows a strong site-dependent variation of the empty-state LDOS. The peaks at + 0.45 V, + 1.05 V and + 1.45 V appear to be most prominent when measured on the 'down atoms' of the Sn tetramer (atom 2 of the 2√3 domain), while the small peak at +1.35 V appears to be most pronounced when measured on the 'up atom' (atom 1). The peak at +0.7 V is mostly associated with atoms in the first layer of Sn (atom 3), away from the tetramer atoms. The LDOS amplitudes of the 4√3Sn structure are much less site specific with the exception of the peak at +1.2 V: it appears to be associated with tunneling into the down atoms (atom 5). This is more clearly seen in the dI/dV spectra, i.e., prior to normalization (see Appendix). Note that the peak at +1.7 V is absent for location 5 in Fig. 12 (b), i.e., the down atom location. This position is registry aligned with the up-atom location of the 2√3 structure, meaning it is located above the 2nd layer Si atom.

We have calculated the LDOS of the 2√3Sn and 4√3Sn structure for both the Törnevik and Revised Trimer model, both with and without boron at the $S_5$ lattice location. The results for the 2√3Sn structures are shown in Fig. 13. One of the most pertinent observations is that the theoretical DOS of the undoped systems and boron-doped systems are very different while no such difference is seen in experiment. A reasonable interpretation of this discrepancy is that the boron atoms of the B-√3 template surface move away from the $S_5$ lattice location during the deposition and annealing of the Sn layer. Since the experimental LDOS of the n-2√3Sn and B-2√3Sn surfaces seem to be identical, at least at 77 K, the boron atoms probably diffused into the bulk. This would be consistent with the fact that the Rp factor in the LEED I-V fitting increased significantly when placing boron at the $S_5$ lattice location. Hence, the present results suggests that the atomic arrangement of the subsurface layers differs from that of the pristine B√3 substrate. For this reason, we will focus on the line-up between the theoretical LDOS of the undoped 2√3Sn and 4√3Sn structures and the normalized dI/dV spectra of the B-2√3Sn and B-4√3Sn surfaces, measured at 4.4 K (n-type data are not accessible at 4.4 K).

Attempts to line up the theoretical DOS of the revised trimer model in Fig. 13(c) with the experimental LDOS presented in Fig. 11 were unsuccessful. In particular, the empty state spectra align rather poorly and one would have to assume that HSE underestimates the gap here. On the other hand, the LDOS of the 2√3Sn Törnevik model in Fig. 13(a) lines up reasonably well, after correcting for the difference in band gap. This is shown in Fig. 14(a). Here HSE seems to overestimate the gap. The agreement is almost perfect when comparing the 2√3Sn Törnevik DOS with the 4√3Sn spectra, as shown in Fig. 14(b). While this comparison may appear odd, keep in mind that experimentally, the peak positions of the 2√3Sn and 4√3Sn spectra in Fig. 11 also line up nicely after correcting for the band gap difference. We finally notice that the HSE LDOS calculation of the Törnevik 4√3Sn structure (not shown) does not align well with the experimental result for either the B-2√3Sn or B-4√3Sn surface. Hence, the 4√3Sn structure from DFT can most likely be disqualified, as we already suspected in Section VI.

### VIII    DISCUSSION AND SUMMARY

On the basis of DFT total energy calculations, the Törnevik model, revised Törnevik model, and the revised trimer model are viable candidates for the 2√3Sn structure. Their relative stability depends on the precise value of the Sn chemical potential, which in turn depends on the experimental growth conditions. Because STM images indicate the presence of a double layer structure, we rule out the revised Törnevik model, the trimer model, and Ichikawa's model. Their simulated STM images also do not reproduce the observed contrast between the up atoms and down atoms within each tetramer. The revised Törnevik model also has a relatively poor Pendry Rp factors in LEED I(V). While the Törnevik and revised trimer models are about equally favored according to our LEED I(V) studies for the n-type samples, the agreement between the theoretical DOS from the DFT-HSE calculations and experimental LDOS from the normalized dI/dV spectra favor the Törnevik model. We therefore conclude that among the five different models considered in this paper, the Törnevik model best agrees with the experimental results.

A number of problems remain. The relatively high Rp factors in LEED I(V) suggest that there is room for further structure optimization. The fact that the revised trimer model appears favored under the experimentally realized chemical potential is somewhat troubling, as is the fact that the DFT calculations have not been able to identify a $4\sqrt{3}$Sn structure as the lowest energy configuration. These uncertainties make it difficult to identify the nature of the $2\sqrt{3}$Sn to $4\sqrt{3}$Sn transition. However, on the basis of the experimental data presented in Fig. 12, we make the following conjecture.

Up-down buckling on semiconductor surfaces is generally accompanied by charge transfer. Here, the dangling bond orbital on the 'up atom' usually acquires additional charge while the one on the 'down atom' gives up some of its charge [45]. Now, the lowest unoccupied state of the $2\sqrt{3}$Sn structure has the greatest amplitude on the down atom, see Fig. 12(a), and the $2\sqrt{3}$ to $4\sqrt{3}$ conversion only happens as electrons start tunneling into this 'anti-bonding' orbital. If the tunneled electrons cannot be drained fast enough, the transient filling of this orbital will 'pull' the down atom upward. This will destabilize the $2\sqrt{3}$ $\begin{smallmatrix} up & up \\ down & down \end{smallmatrix}$ buckling geometry of the $2\sqrt{3}$ structure, which then would rationalize a displacive transition to a $4\sqrt{3}$ structure with the $\begin{smallmatrix} up & down \\ down & up \end{smallmatrix}$ buckling geometry. According to the $4\sqrt{3}$ spectra in Fig. 12(b), the peak at +1.2 eV now has the largest amplitude on the down atoms of the $4\sqrt{3}$Sn structure and one would expect this geometry to become unstable when tunneling into the +1.2 eV state of the $4\sqrt{3}$Sn structure. Indeed, this is what happens. Fig. 15 shows a series of images showing the appearance of the $4\sqrt{3}$Sn structure when tunneling into the lowest unoccupied state of the $2\sqrt{3}$Sn structure and the reversal of this process for tunneling bias greater than 1.25 eV.

At 4.4 K, the $2\sqrt{3}$ to $4\sqrt{3}$ conversion likely involves quantum tunneling of the Sn adatoms through a tunneling barrier [46, 47]. The barrier is likely asymmetric as the $4\sqrt{3}$ structure appears lowest in energy, at least for the p-type system (Fig. 16). While we do not know the geometry of the transition state in between the 2√3Sn and 4√3Sn structures, it likely has a reduced band gap or it may even be metallic, as its total energy should be higher [47]. It possibly represents a flat tetramer configuration [17].

The experimental observations suggest that the transient filling of the 2√3 down-atom dangling-bond orbital destabilizes the 2√3 structure relative to the transition state, i.e., it reduces the activation barrier, resulting in a local 2√3 to 4√3 conversion. Likewise, transient filling of the 4√3 down-atom dangling-bond orbitals destabilizes the 4√3 structure relative to the transition state, resulting in a 4√3 to 2√3 conversion. These conversions happen at different bias, as shown in Fig. 15.

From the fact that the transition is only seen on the p-type substrates, we infer that the total energy of the 2√3 and 4√3 structures (and the energy barrier between the two) also depends on the fixed doping level and dopant type (Fig. 16). Indeed, our DFT calculations for the 4√3 structures identified in Section VI indicate that hole doping reduces the total-energy of the 4√3 phase relative to the 2√3 phase by as much as 150 meV per 2√3 unit cell [48]. Note, however, that the 2√3 structure is still the most stable structure in these calculations, so the precise role of the dopant atoms remains unclear at this point.

Summarizing, the experiments indicate that both fixed and transient doping destabilizes the 2√3Sn and 4√3Sn states relative to the transition state, and facilitates the transition between the two states. Clearly, better understanding of the thermodynamic and kinetic aspects of the transition requires a firm resolution of the 4√3Sn ground state structure and identification of a transition pathway. It also requires a better understanding of the excited state decay and carrier dynamics under the tunneling conditions. Nonetheless, the arguments presented here are rather generic and hence, they should be more broadly applicable. As such, hole doping appears a viable strategy toward the creation of novel surface phases with potentially interesting and desirable electronic properties.

**Acknowledgement.** This work was primarily funded by the National Science Foundation under grant No. DMR 1410265.    J.H.C. is supported by the National Research Foundation of Korea (NRF) grant, funded by the Korea Government (Grants No. 2015R1A2A2A01003248 and No. 2015M3D1A1070639).    The calculations were performed at the KISTI supercomputing center through the strategic support program (KSC-2016-C3-0059) for the supercomputing application research.

**Table I**. Ratio of normalized core level peaks for different Sn core levels and escape angles, along with the calculated coverage (in monolayers) from a layer-by-layer attenuation model (see Appendix).

| θ | $I_{2\sqrt{3}} / I_{\sqrt{3}}$ | | | Coverage (ML) | | |
|---|---|---|---|---|---|---|
| | Sn 3p | Sn 3d | Sn 4d | Sn 3p | Sn 3d | Sn 4d |
| 52° | 3.49 | 4.09 | 3.70 | 1.08 | 1.27 | 1.14 |
| 27° | 3.34 | 3.39 | 2.84 | 1.04 | 1.05 | 0.88 |
| 12° | 3.22 | 3.80 | 2.77 | 1.00 | 1.18 | 0.86 |

**Table II.** Pendry R factor for different models after the vertical relaxation, Debye temperature optimization and curve smoothing.

| Pendry R factors | Tornevik's model | Revised Tornevik's model | Trimer model | Revised trimer model |
|---|---|---|---|---|
| n-2√3 | 0.34 | 0.39 | 0.30 | 0.35 |
| B-2√3 | 0.38 | 0.41 | 0.38 | 0.43 |

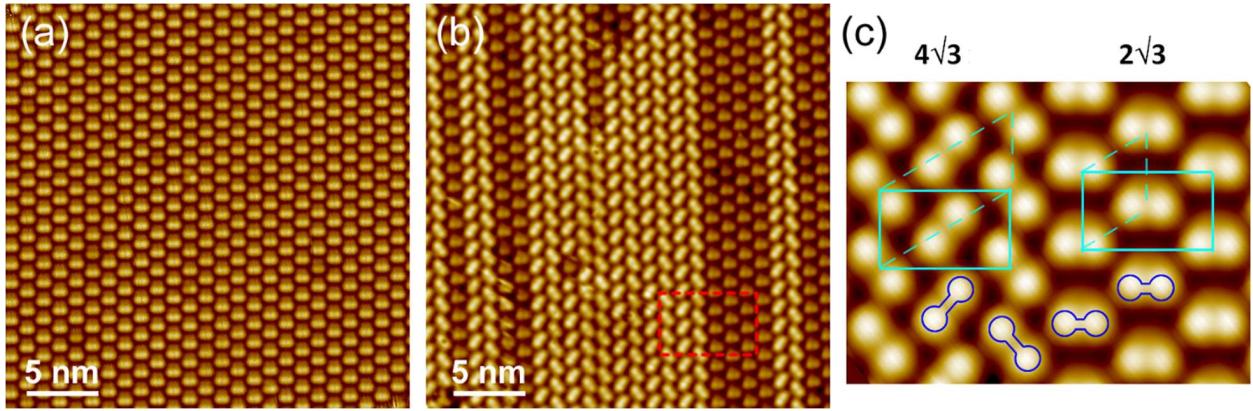

**Figure 1.** Scanning Tunneling Microcopy images of the hole doped Si(111)(2√3×2√3)R30º-Sn interface. Panel **(a)** shows a uniform 2√3 phase at room temperature. The image in panel **(b)** shows coexisting 2√3 and 4√3 domains at 4.4 K. The 2√3 to 4√3 transition does not occur in the electron-doped system (images not shown; see Ref. 17). The tunneling parameters are +2V and 0.1 nA in panels **(a)** and +1.5 V, 0.02 nA in **(b)** . Panel **(c)** shows a close up image (+0.8 V, 0.15 nA) of the surface area marked in (**b**). Dashed lines mark the primitive (2√3×2√3)*R*30º and (4√3×2√3)*R*30º unit cells of the 2√3Sn and 4√3Sn phase, respectively, while solid lines mark the corresponding centered-rectangular and rectangular unit cells. Up-dimers are marked by dumbells.

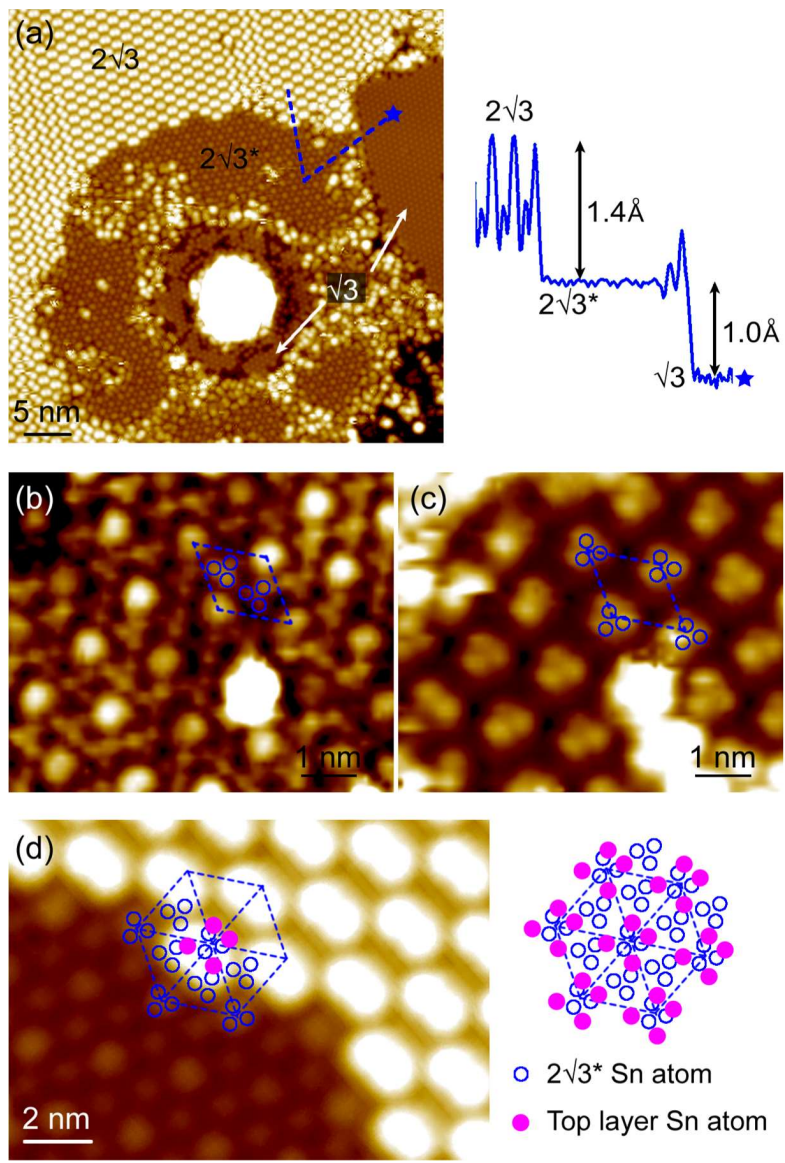

**Figure 2.** **(a)** STM image (+0.8V, 0.03 nA) of the B2√3-Sn surface after applying a +2 V voltage pulse to the STM tip. The large circular area was completely covered with the 2√3-Sn surface before the pulse was applied (image not shown), except for the presence of the native √3-Sn area on the far right of the image. Before pulsing the tip, the tip-sample separation was decreased by about 0.5 nm relative to the normal imaging set point. The bright area below the center of the image likely represents a pile of Sn atoms

originating from the top layer of the 2√3 reconstruction. Just outside the bright area is a ring of a newly created √3-Sn structure and an unknown 2√3* phase. The latter is the exposed underlayer of the 2√3-Sn surface. It also has a (2√3×2√3)$R$30° periodicity but the structure is different from the 2√3 reconstruction. The 2√3* layer is higher than the neighboring √3 structure but lower than the neighboring 2√3 structure, indicating that the 2√3* structure is a single layer structure but with a higher Sn atom density as compared to the √3 phase. The line scan illustrates the bilayer nature of the original 2√3 reconstruction. Outside the circular 2√3* region, the native 2√3 phase is coexisting with 4√3 phase. The latter must be induced by the positively biased pulse or during the scanning process. Panels **(b)** and **(c)** are STM images of the same area of the 2√3* structure, scanned at different tunneling conditions (+0.1 V 0.02nA for **(b)**; +1.3 V 0.02 nA for **(c)**), revealing the existence of two downward-pointing trimers and one upward-pointing trimer, indicating a coverage of nine Sn atoms per 2√3 unit cell, or 9/12 ML.    Panel **(d)** illustrates the registry of the top layer Sn atoms of the neighboring 2√3 domain on top of the 2√3* underlayer structure. This registry is used as a starting point for the structure refinement in DFT and LEED.

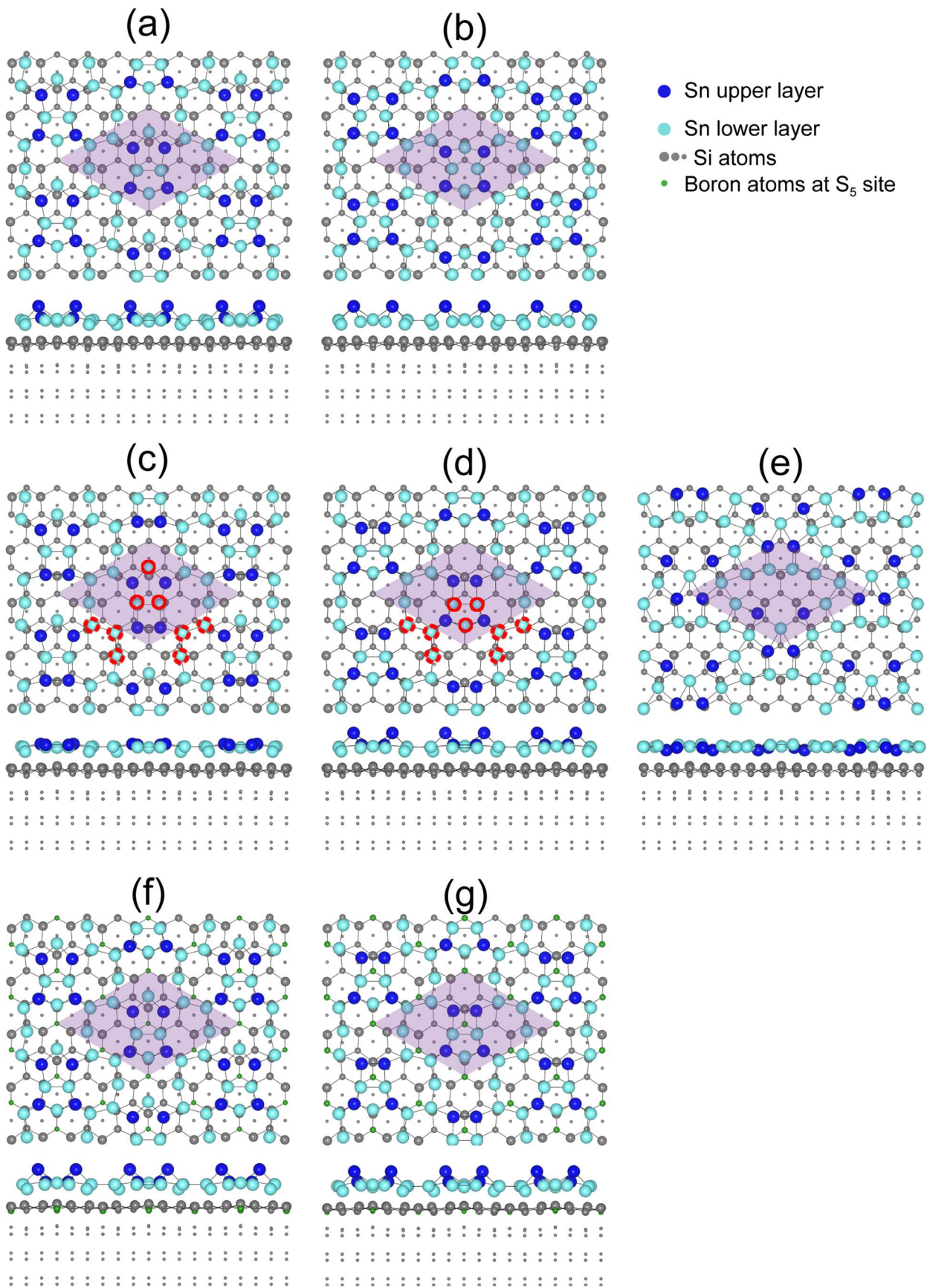

(a)        (b)

● Sn upper layer
● Sn lower layer
● Si atoms
● Boron atoms at $S_5$ site

(c)        (d)        (e)

(f)        (g)

**Figure 3**. Top view and side view of the total-energy-minimized geometry of the 2√3-Sn structure for (a) the Törnevik model [18]; (b) the revised Törnevik model; (c) Trimer model; (d) Revised trimer model; and (e) Ichikawa model [21]. The encircled Sn atoms in panels (c) and (d) are the original trimer atoms of the 2√3* structure prior to the DFT structure optimization of the 2√3Sn structure (see text). Panels (f) and (g) show the total energy minimized geometry of the Tornevik and revised trimer model, respectively, in the presence of a boron underlayer. The boron atoms are placed at the $S_5$ lattice locations [27, 28] in a way that preserves the *cm* space group symmetry. There are four boron atoms per 2√3 unit cell. The structures in (a), (b), and (f) have a total Sn coverage of 14/12 ML. Those in panels (c), (d), (e), and (g) have a coverage of 13/12 ML.

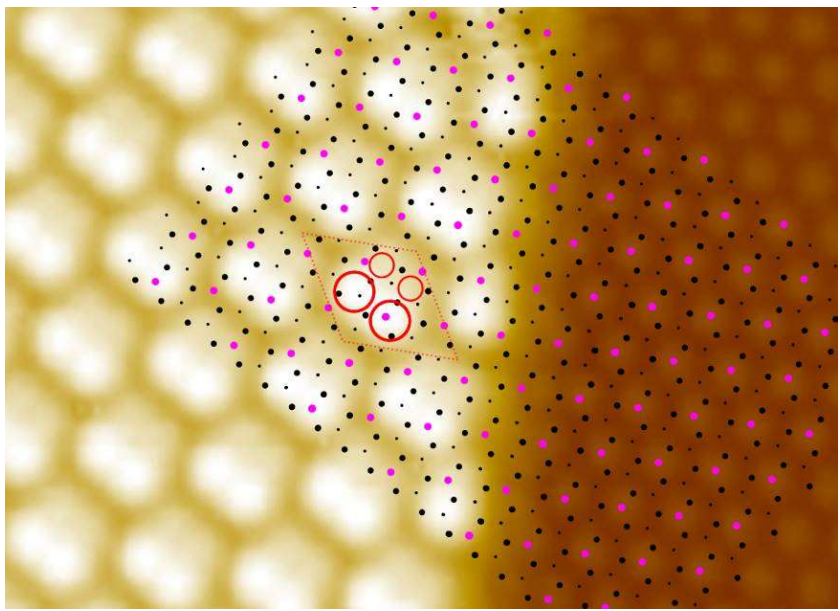

**Figure 4**. STM image showing coexisting domains of the $\sqrt{3}$ and $2\sqrt{3}$ reconstructions. The $\sqrt{3}$ reconstruction consists of 1/3 ML of Sn adsorbed at the $T_4$ adatom sites of the Si(111) substrate. Note that there are three $T_4$ sites per $\sqrt{3}$ unit cell. The grid overlay thus allows us to determine the registry of the top layer atoms of the $2\sqrt{3}$ reconstruction (see text). The black solid dots are the atoms of the outermost Si bilayer and the purple solid dots mark the $T_4$ adsorption sites. The rhombic $2\sqrt{3}$-Sn unit cell is indicated. The 'up' and 'down' atoms of the Sn tetramer are indicated by the large and small red circles, respectively. The up atoms are approximately located at the $T_4$ sites while the down atoms are located above the three fold hollow ($H_3$) sites of the Si(111) substrate.

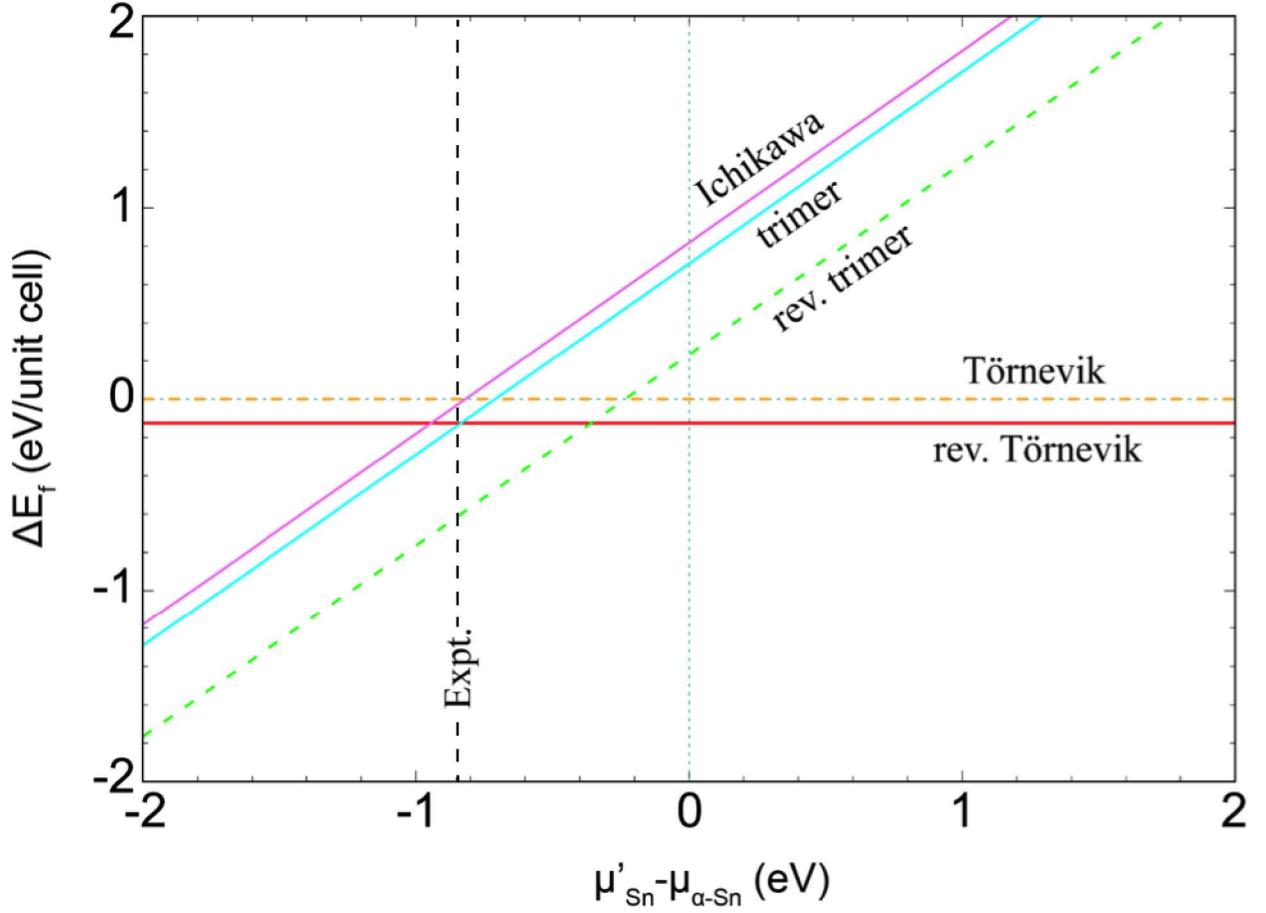

**Figure 5.** Formation energies of the candidate structures presented in Fig. 3, calculated as a function of the chemical potential of the Sn adatoms, according to Eq. 1. The chemical potential is determined by the experimental growth conditions and is estimated using Eq. 2 to be -0.85 eV (dashed vertical line).

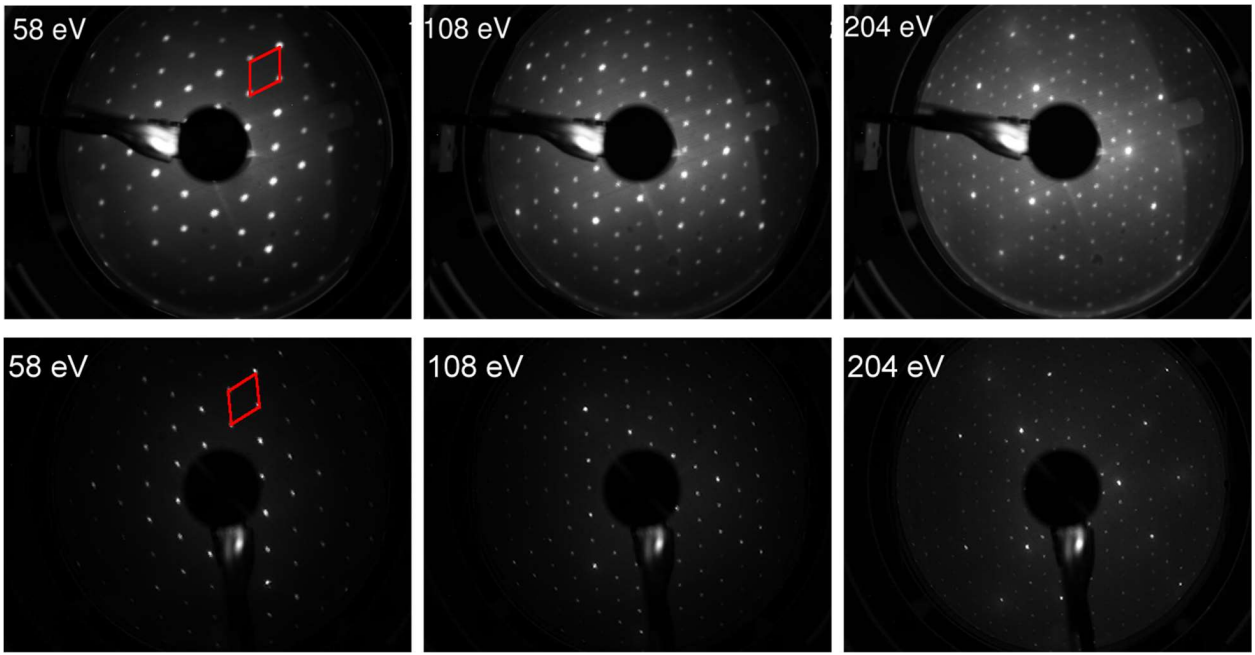

**Figure 6.** LEED patterns of the B-2√3Sn (top) and n-2√3Sn (bottom) surfaces, recorded at three different beam energies and acquired at a sample temperature of 90 K. The unit cell of the reciprocal lattice is indicated in red in the two panels on the left (58 eV).

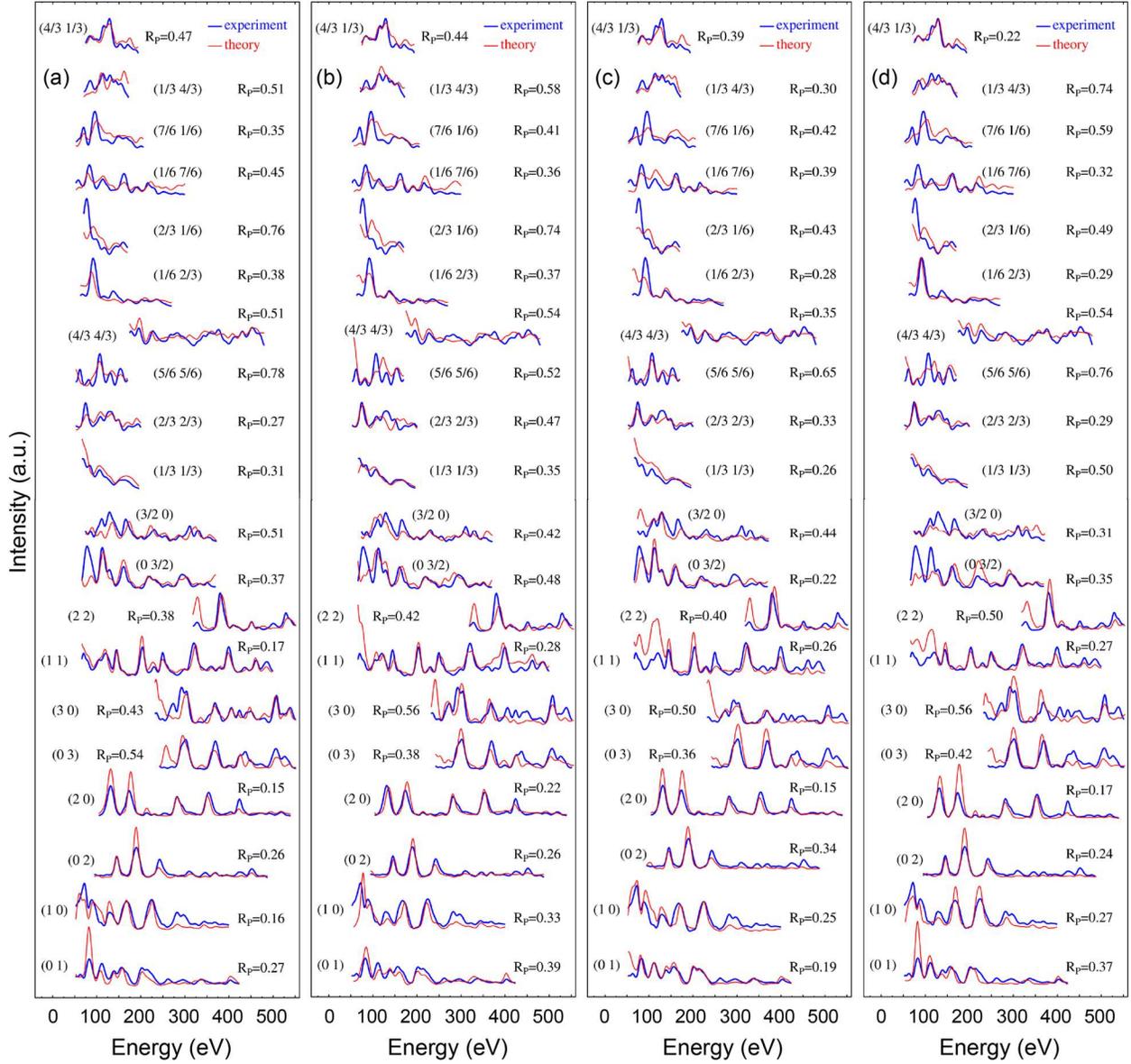

**Figure 7.** LEED I-V data of the n-2√3Sn surface, and best fit calculated intensities according to (a) the Törnevik model; (b) revised Törnevik model; (c) trimer model; and (d) revised trimer model. R-factors for the individual beams are indicated. The overall R-factors for these models are listed in Table 2.

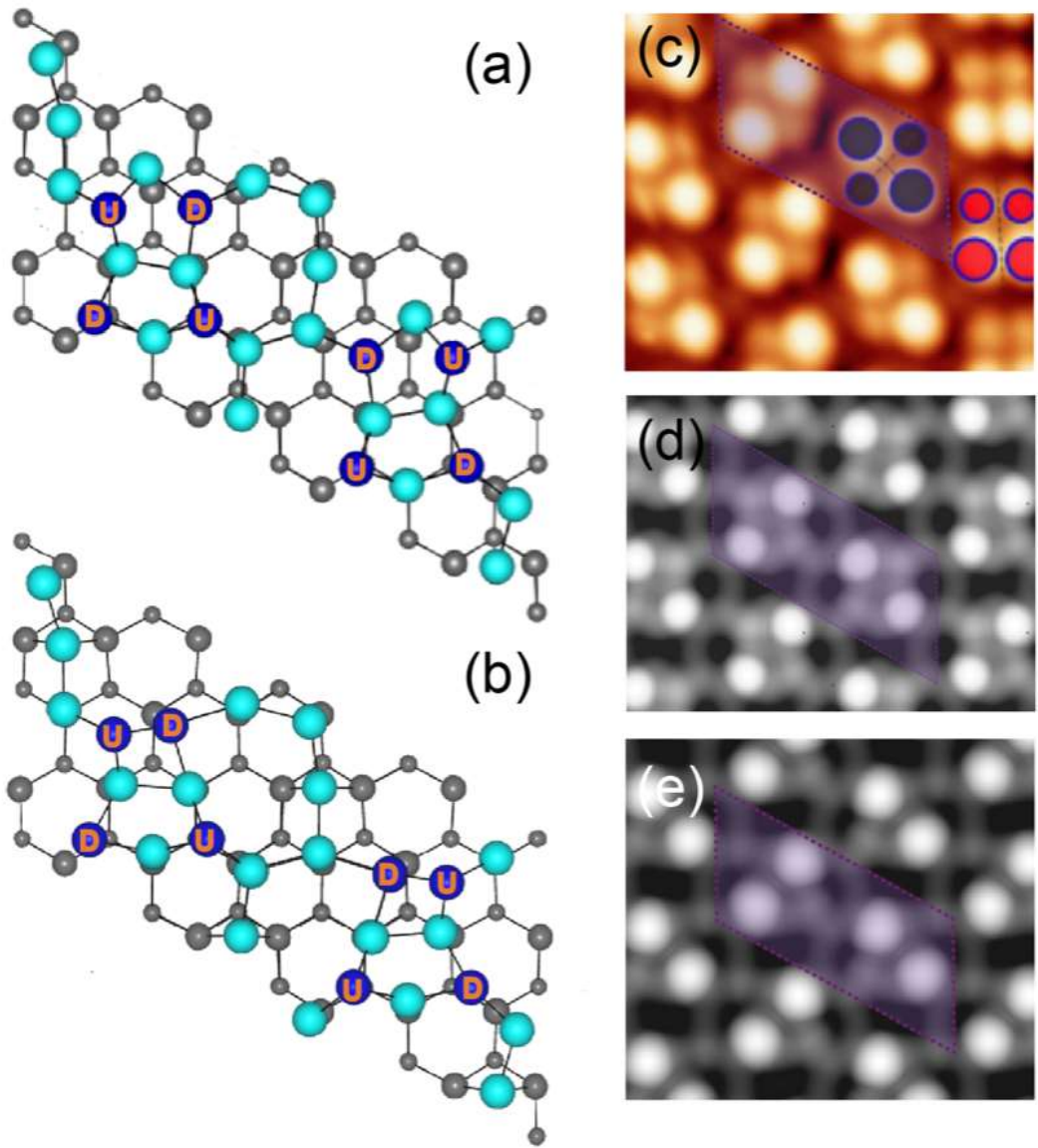

**Figure 8.** Fully relaxed geometry of the 4√3 structure, starting with the optimized Törnevik coordinates (a) and revised trimer (b) coordinates. The up (U) and down (D) atoms of the Sn tetramers are indicated. (c) Empty state experimental STM image of the B-4√3Sn structure, scanned at +1 V, 0.2 nA. This STM image is noise filtered. Note the existence of a narrow 2√3 strip on the far right of the image. The up atom and down atom positions are indicated by large and small filled circles, respectively (dark blue for the 4√3 structure and red for the 2√3 structure). Simulated empty state STM images of the 4√3 structures based on the Törnevik (d) and revised trimer (e) models, all at + 1V bias voltage. The purple shaded parallelograms mark the $(4\sqrt{3}\times2\sqrt{3})R30°$ unit cell.

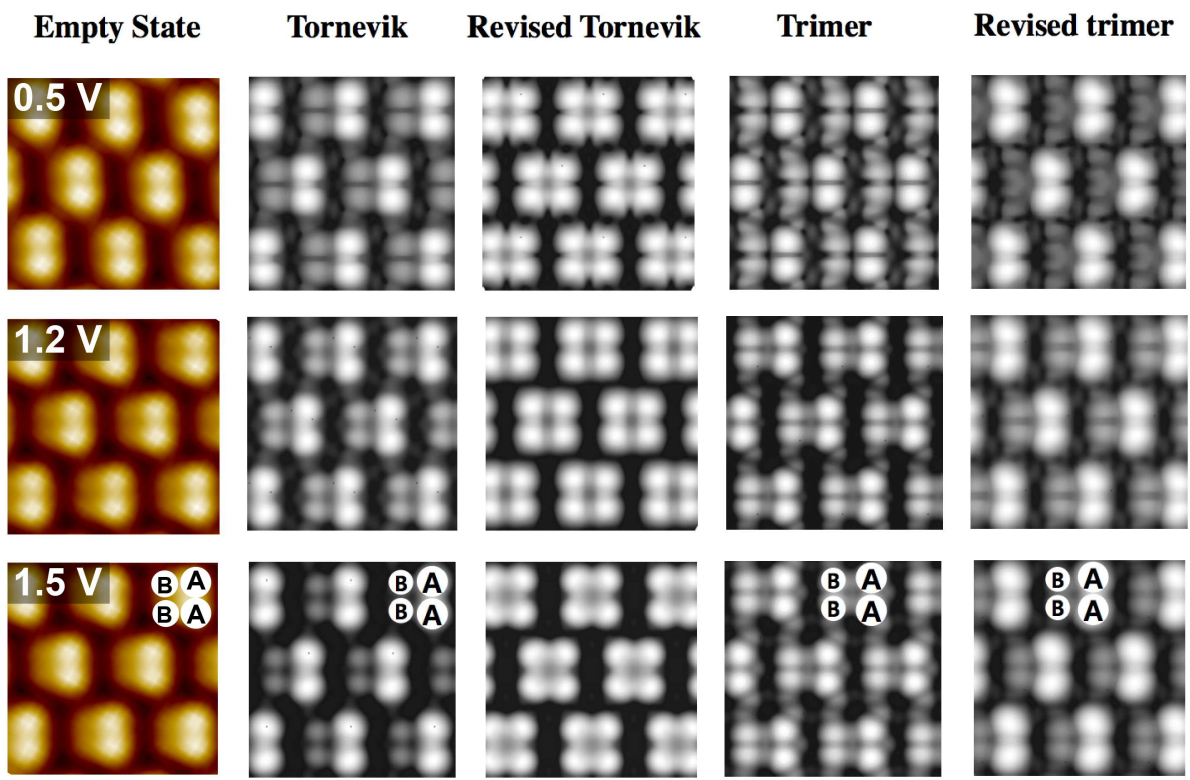

**Figure 9.** Experimental (left column) and simulated grayscale STM images for various structure models. The experimental and theoretical images are compared for three different empty state biases, as the lower dimer atoms are not well resolved in filled state images.

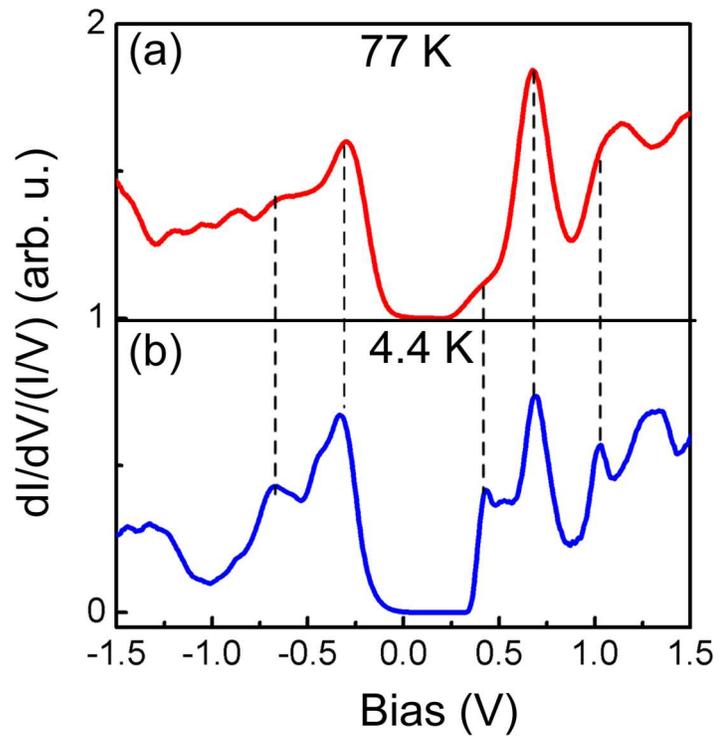

**Figure 10.** Line-up of the normalized dI/dV spectra of the B-2√3Sn surface, recorded at (a) 77 K and (b) 4.4 K. The 77 K spectrum was published in Ref. 17. Both spectra were recorded with the STM tip positioned above an up-atom of the B-2√3Sn structure.

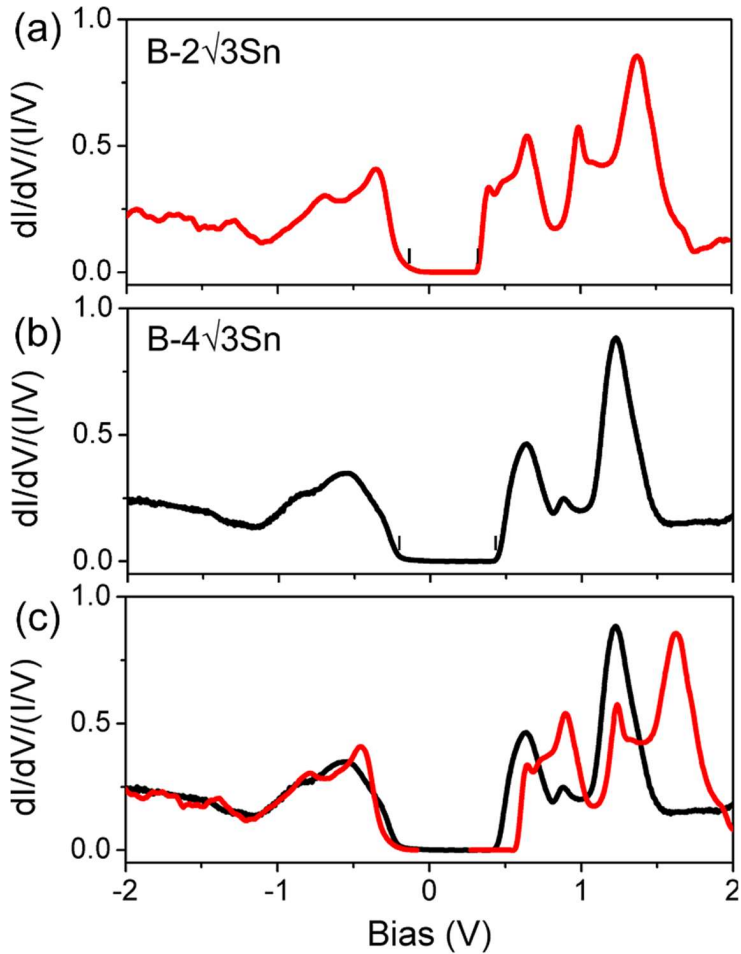

**Figure 11**. Normalized dI/dV spectra from STS recorded of the (a) B-2√3Sn and (b) B-4√3Sn surface, measured at 4.4 K. Both spectra were recorded with the STM tip positioned above a down-atom. Panel (c) shows a line-up of the spectra after stretching the band gap in (a) (the filled states shift 0.1 V toward lower energy, while the empty states shift 0.24 V toward higher energy) to match the band gap value in (b). Note the absence of a peak at +1.7 V in the B-4√3Sn spectrum. Such a peak does exist, however, when the spectrum is recorded on the up-atom (Fig. 12). The band gap values in (a) and (b) are 0.45 and 0.70 eV (marked by tick marks), respectively.

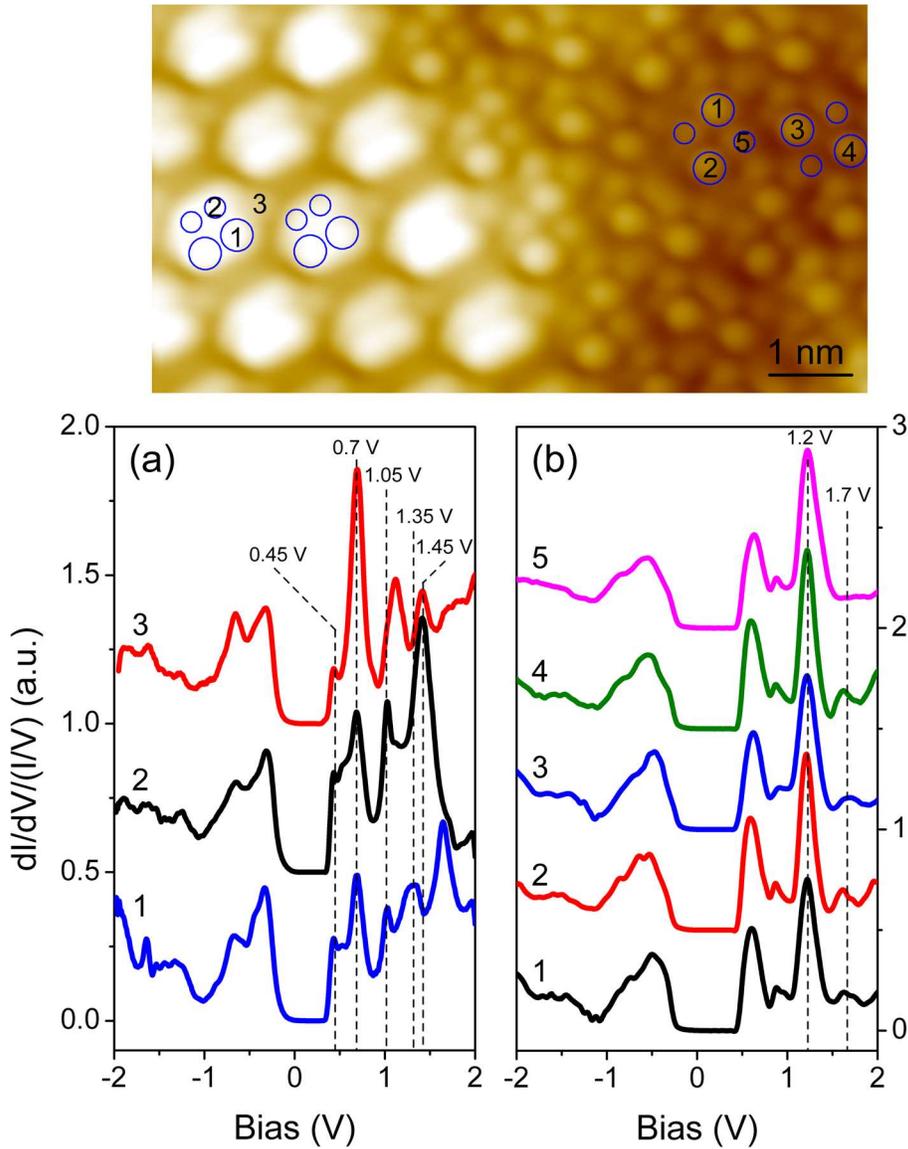

**Figure 12**. Normalized dI/dV spectra, recorded at specific atomic locations within neighboring B-2√3Sn (left) and B-4√3Sn (right) domains. The spectra are labeled according to the numbered site locations in the top panel. This STM image was scanned at -0.01V and 50 pA. For a detailed explanation, see text. While the spectra in panel (b) are all very similar, the corresponding dI/dV spectra in the Appendix clearly indicate that the +1.2 V peak is strongly localized on atom 5, i.e., the down atom.

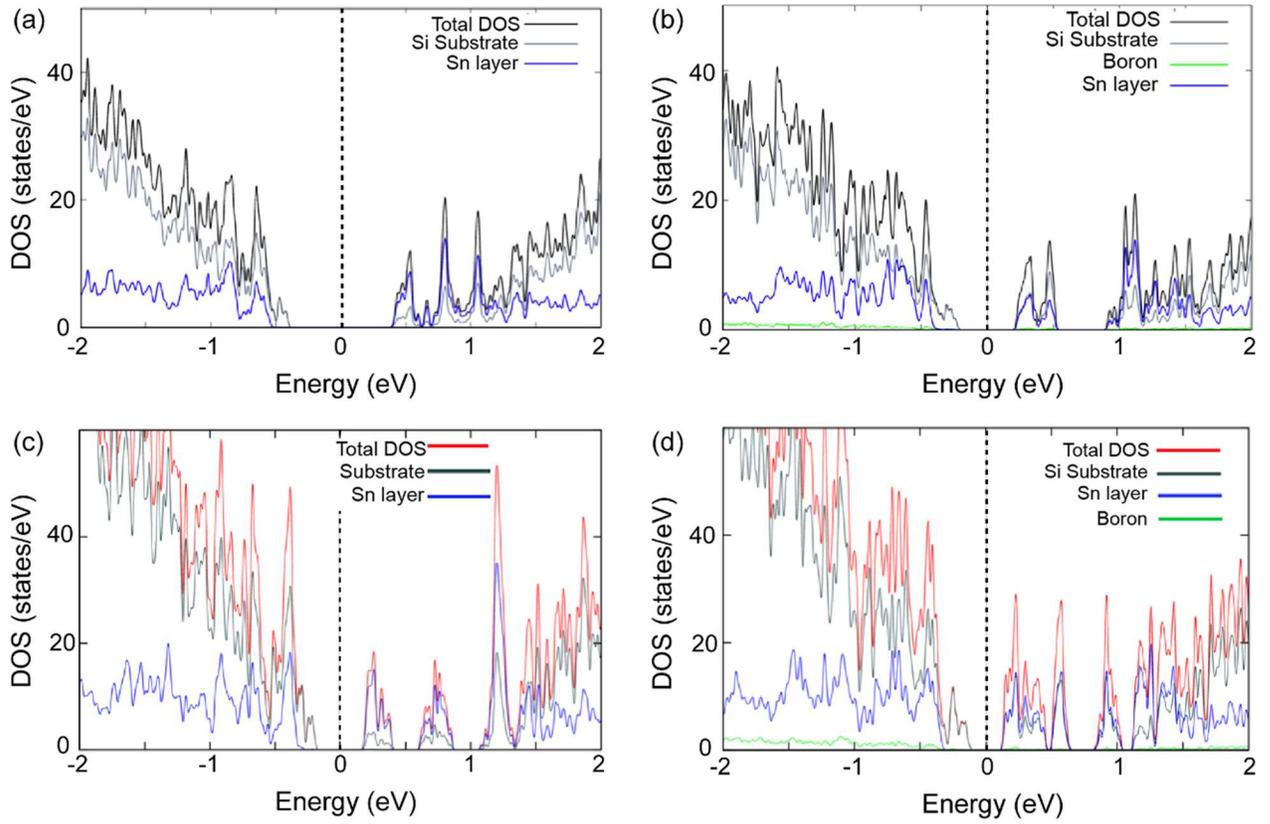

**Figure 13.** DFT calculations of the density of states using the HSE functional. Panels (a) and (b) show the density of states for the 2√3 Törnevik model, without and with a (√3×√3)R30º boron underlayer, respectively. Panels (c) and (d) show the density of states for the 2√3 revised trimer model without and with the (√3×√3)R30º boron underlayer, respectively.

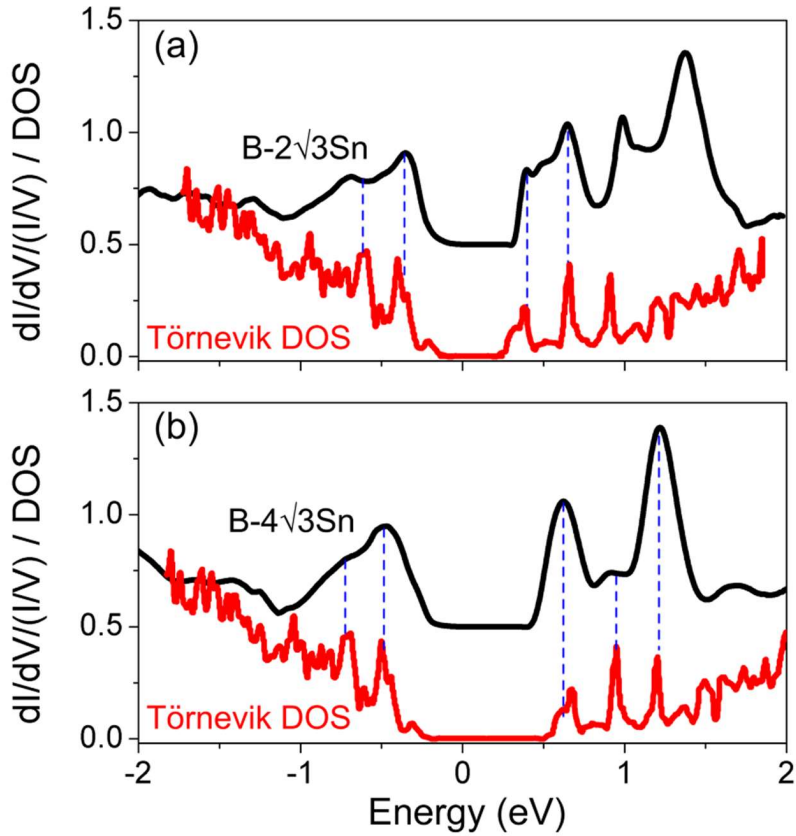

**Figure 14.** (a) Line up of the normalized dI/dV spectrum of the B-2√3Sn surface and the corresponding density of states according to the undoped Törnevik model. The theoretical band gap has been adjusted (spectrum below $E_F$ shifted 0.25 eV toward higher energy; spectrum above $E_F$ shifted 0.15 eV toward lower energy) so as to match the experimental result. (b) same as in (a) but with the experimental spectrum replaced by the normalized dI/dV spectrum of the B-4√3Sn surface, and showing an improved line-up. Here, the full theoretical spectrum was shifted 0.15 eV toward higher energy without the artificial band gap correction.

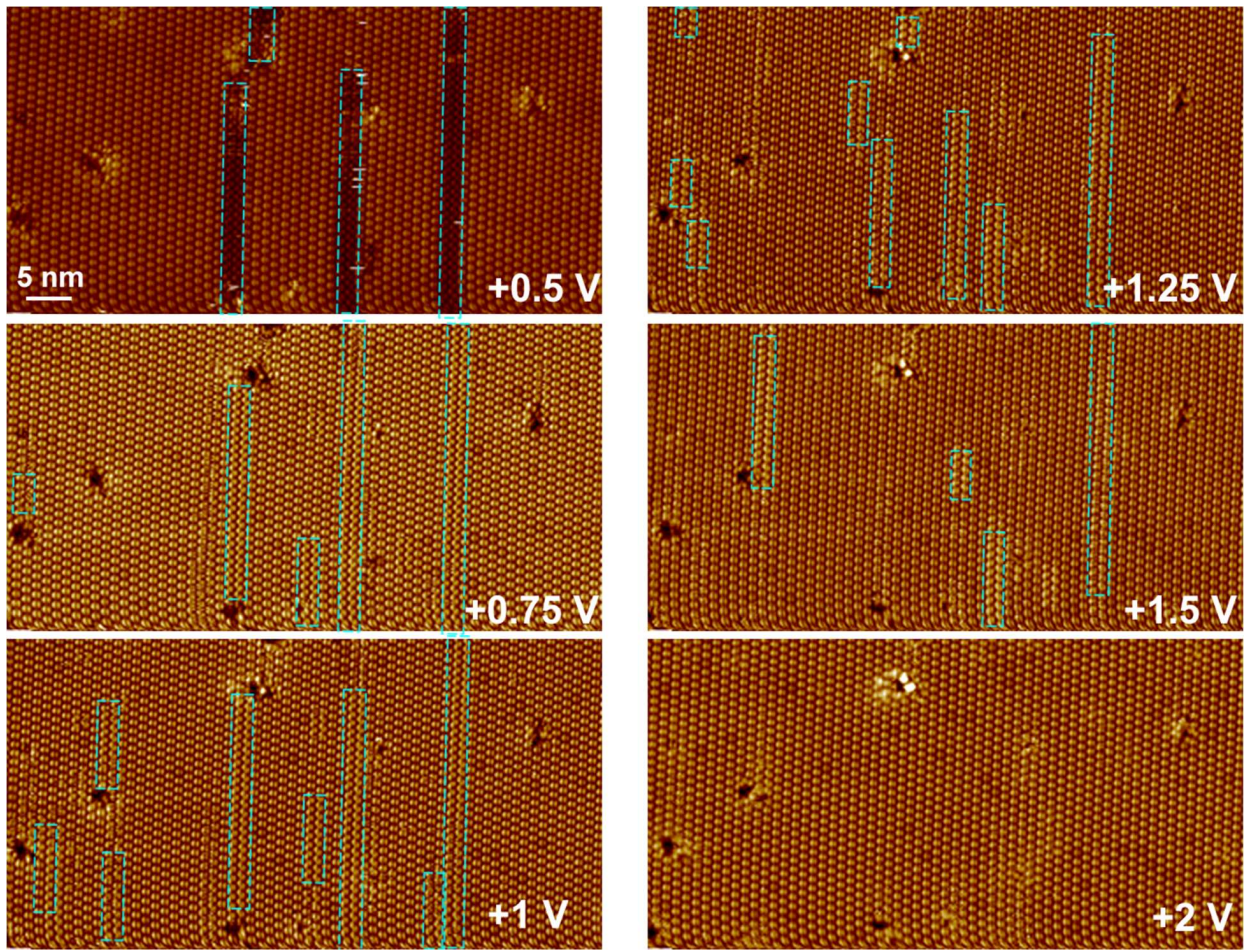

**Figure 15.** Bias dependent STM images of the p-2√3 phase at 77 K. The images are taken on the same area, and with the same tunneling current of 1 nA. The 4√3 domains are marked with dashed rectangles. It can be seen that the 4√3 fraction reaches at maximum at a scanning bias between 1.00 and 1.25 V.

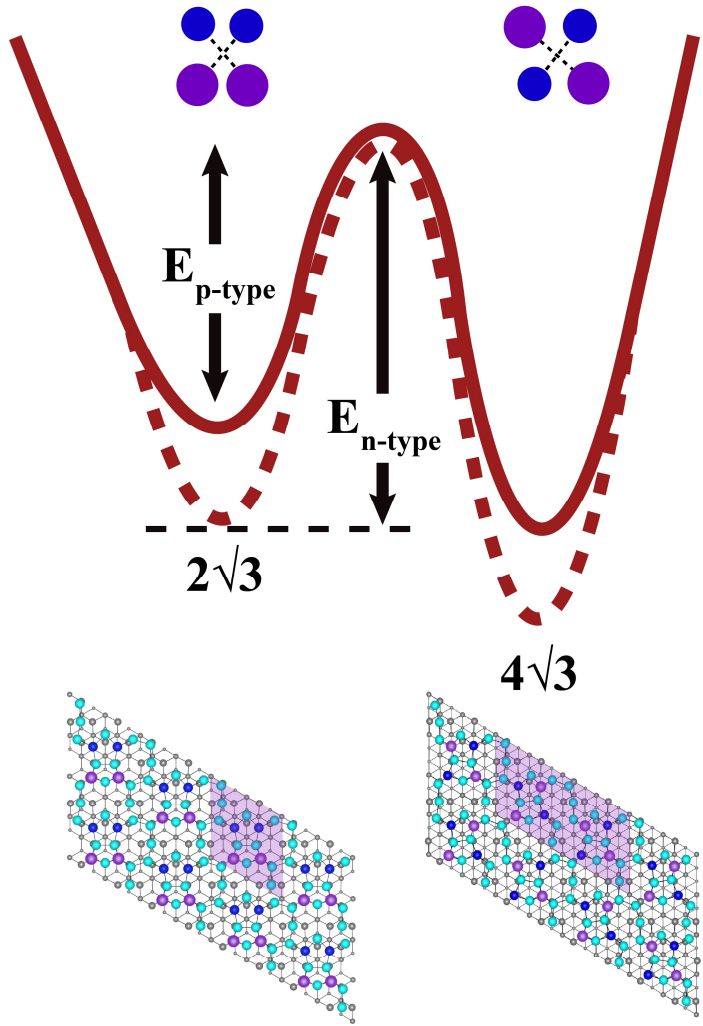

**Figure 16.** A possible transition state diagram for the quantum tunneling transition between the $2\sqrt{3}$ and $4\sqrt{3}$ structures, where the $4\sqrt{3}$ structure is assumed to be the most stable configuration for both the n- and p-type samples. The energy diagrams of the electron-doped and hole-doped systems are indicated by the dashed and solid lines, respectively. In this particular case, the barrier for the $2\sqrt{3}$ to $4\sqrt{3}$ conversion is lowest for the p-type system, while the reverse transition remains highly unlikely for both the n-type and the p-type systems. The total-energy minimized structures of the $2\sqrt{3}$Sn (left) and $4\sqrt{3}$Sn (right) phases are shown for easy comparison.

# Appendix I

## Coverage estimate for the 2√3-Sn surface from XPS

The intensity of the Sn core level spectrum of a monatomic Sn layer is proportional to the number of Sn atoms in the layer. Hence, it should be possible to determine the absolute coverage of the 2√3 surface by comparing the Sn core level intensity of the 2√3 surface to that of the √3 surface, whose coverage is known to be exactly 1/3 ML. In practice, however, this method can be quite imprecise because it requires perfect reproducibility of the sample alignment and other experimental parameters when collecting XPS spectra from the two samples. To eliminate these experimental uncertainties, we normalize the Sn 3$p$, 3$d$, and 4$d$ core level intensities to the Si 2$p$ core level intensity measured on the same sample.

When comparing the normalized core level intensities of the 2√3 and √3 surfaces, one must take into account that the Si 2$p$ core level intensity is attenuated due to both elastic and inelastic scattering of the Si 2$p$ photoelectrons when passing through the Sn layer [49]. The attenuation factors differ for the 2√3 and √3 surfaces due to the different adatom densities in the respective Sn layers. In this note, we describe how the different attenuation factors are taken into account, and show that inclusion of these attenuation factors lead to coverage corrections of less than 10%.

If the number of Sn adatoms in the √3 and 2√3 structures is given by $N_1$ and $N_2$, respectively, we can write the corresponding core level intensities as follows:

$$I_{Sn}^{\sqrt{3}} = \alpha N_1 \tag{A1}$$

$$I_{Sn}^{2\sqrt{3}} = \alpha N_2 \tag{A2}$$

where $\alpha$ is a proportionality constant, which depends for instance on the photo-ionization cross-section, sample position, x-ray flux, and transmission characteristics of the electron energy analyzer. Next, it is assumed that the Si 2$p$ core level intensity from the substrate $I_{Si}^{0}$ is attenuated by a factor $\exp(-d/\lambda \cos \vartheta)$ following Beer-Lambert's law, where $d$ is thickness of the Sn layer and $\lambda$ the effective attenuation length of the Si 2$p$ photoelectron [49]. The angle $\vartheta$ is the angle between the photoelectron detector and the surface normal. Accordingly,

$$I_{Si}^{\sqrt{3}} = \frac{2}{3} I_{Si}^{0} + \frac{1}{3} I_{Si}^{0} \exp(-d/\lambda \cos \vartheta) \tag{A3}$$

$$I_{Si}^{2\sqrt{3}} = I_{Si}^0 \exp\left(-d/\lambda\cos\vartheta\right) \tag{A4}$$

Here, we assume that the Si 2$p$ core-level attenuation factor of the 2$\sqrt{3}$ surface equals that of a Sn layer with coverage of 1.0 ML, whereas in effect, the absolute coverage has yet to be determined. In light of the fact that the absolute coverage is known to be close to 1 ML, and that the total correction factor for the attenuation of the Si 2$p$ photoelectrons is less than 10%, this subtle point only introduces a very small error. Taking the ratio $R$ of the normalized intensities (Table 1), we obtain

$$R = \frac{I_{Sn}^{2\sqrt{3}}/I_{Si}^{2\sqrt{3}}}{I_{Sn}^{\sqrt{3}}/I_{Si}^{\sqrt{3}}} = \frac{N_2}{N_1} \times \frac{2 + \exp\left(-d/\lambda\cos\vartheta\right)}{3\exp\left(-d/\lambda\cos\vartheta\right)} \tag{A5}$$

From which the ratio $N_1/N_2$ and hence the absolute coverage of the 2$\sqrt{3}$ structure can be determined.

Taking into account that, according to our STM results, 10.6 ± 2.4 % of the adatom sites in the $\sqrt{3}$Sn structure are occupied by substitutional Si defects and 7.4 ± 2.2 % of the 2$\sqrt{3}$ sample is covered by the low-density $\sqrt{3}$ structure, equation (A5) can be modified as follows:

$$R = \frac{I_{Sn}^{2\sqrt{3}}/I_{Si}^{2\sqrt{3}}}{I_{Sn}^{\sqrt{3}}/I_{Si}^{\sqrt{3}}} = \frac{\left(N_1 P_1\left(\exp\left(-d/\lambda\cos\vartheta\right)-1\right)+1\right)\times\left(N_2 P_2 + \left(1-P_2\right)N_1\right)}{N_1 P_1 \exp\left(-d/\lambda\cos\vartheta\right)} \tag{A6}$$

where $P_1$ = 0.894 and $P_2$ = 0.926.

The spectra were taken at three different escape angles $\vartheta$ of 52°, 27°, and 12°. The Sn 3$p$, 3$d$, 4$d$, and Si 2$p$ core level peak intensities were obtained by integrating the core level spectra using either a Shirley or linear background, depending on the background conditions. Plasmon-loss satellites were included in the fit. Using Eq. A6 with $d$ = 2.4 Å (Fig. 1) and λ= 26 Å [49], we obtain the absolute coverage of the 2$\sqrt{3}$Sn surface, as shown in Table 1 of the main text. The above value of λ was calculated for α-Sn. Since the atomic density in the 2$\sqrt{3}$ surface is higher than that of an α-Sn(111) layer, λ is likely overestimated. While λ can be corrected for the difference in atomic densities [49], it would hardly affect the outcome of the coverage calculation.

# Appendix II

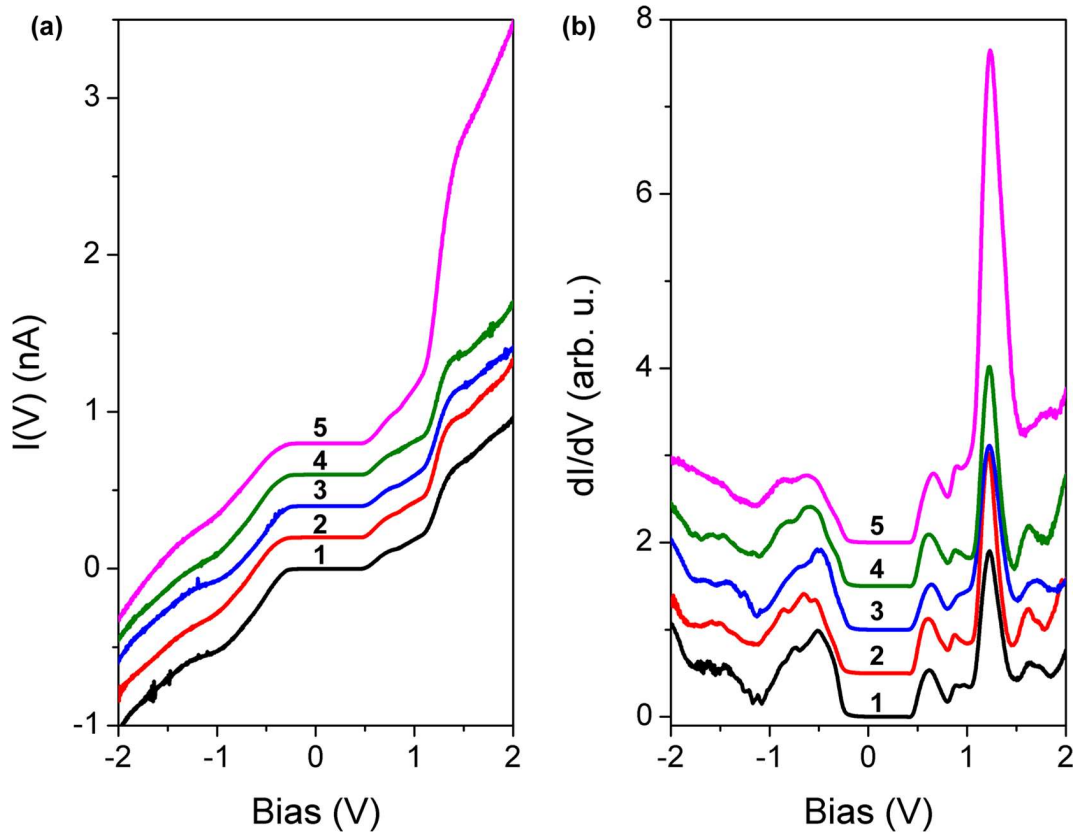

**Figure A1:** (a) Raw I(V) and (b) dI/dV spectra of the B-4√3Sn surface, measured at 4.4 K. The corresponding normalized dI/dV data are shown in Fig. 12(b).

Fig. A1(a) shows a very steep increase in the tunneling current of the B-4√3Sn surface at about +1.2 V, especially for location 5 in Fig. 12, which is the location of the down atom. The corresponding dI/dV spectrum in Fig. A1(b) shows a very intense peak at +1.2 V, indicative of tunneling into a strongly localized state with a high LDOS. In order to obtain a more accurate reflection of the LDOS, it is customary to divide the dI/dV spectrum by the I(V) spectrum, so as to divide out the exponential voltage dependence of the tunneling probability [43]. However, in this particular case, the tunneling current near 1.2 V rises much more steeply than expected on the basis of the tunneling probability alone. The steep rise near 1.2 V implies that dI and I become comparable in magnitude so that $dI/I \rightarrow 1$. Accordingly, when the LDOS is very high, as shown here by the steep rise in tunneling current, the peak intensity in the normalized dI/dV spectrum tends to saturate, resulting in an underestimation of the true LDOS of the corresponding quantum state.